\documentclass[a4paper, 10pt]{article}

\usepackage[english]{babel}
\usepackage{url}
\usepackage{hyperref}
\usepackage{color}
\usepackage[usenames,dvipsnames]{xcolor}
\usepackage{amsmath,amsfonts, slashed, amssymb, wrapfig,subfigure}
\usepackage[latin1,ansinew]{inputenc}
\usepackage{fullpage}
\usepackage[squaren,italian,Gray,thinspace,thickqspace]{SIunits}
\usepackage{graphicx}
\usepackage{graphics}
\usepackage{pdflscape}
\usepackage{verbatim}
\usepackage{multirow}
\usepackage[gen]{eurosym}

\newcommand{\coniglietti}[1]{\textquotedblleft #1\textquotedblright}
\newcommand{\reff}[1]{(\ref{#1})}
\newcommand{\beq}{\begin{equation}}
\newcommand{\eneq}{\end{equation}}
\newcommand{\definition}{\overset{\text{def}}{=}}

\newcommand{\checker}{\genfrac{}{}{-3pt}{3}{\blacksquare\negthinspace\square}{\square\negthinspace\blacksquare}}
\newcommand{\antichecker}{\genfrac{}{}{-3pt}{3}{\square\negthinspace\blacksquare}{\blacksquare\negthinspace\square}}

\title{From innovation to diversification: a simple competitive model}
\author{Fabio Saracco$^{1}$, 
Riccardo Di Clemente$^{1}$ \thanks{\mbox{Corresponding author: \emph{E-mail~address}:~riccardo.diclemente@imtlucca.it}}
\& Andrea Gabrielli$^{1,2}$ \& Luciano Pietronero$^{1,3,4}$\\
\footnotesize
$^{1}$\textit{Istituto dei Sistemi Complessi - CNR, Via dei Taurini 19, 00185 Rome, Italy}\\
\footnotesize
$^{2}$\textit{IMT Institute for Advanced Studies Lucca, Piazza S. Ponziano 6, 55100, Lucca, Italy}\\
\footnotesize
$^{3}$\textit{Sapienza, Universit\`a di Roma, Piazzale A. Moro 5 Roma Italy.}\\
\footnotesize
$^{4}$\textit{London Institute for Mathematical Sciences, 35a South St, Mayfair London United Kingdom.}
}

\begin{document}
\date{Published 06-Nov-2015 on PloS One 10(11): e0140420\\  \href{http://dx.doi.org/10.1371/journal.pone.0140420}{DOI: 10.1371/journal.pone.0140420} (2015)}
\maketitle

\begin{abstract}

\noindent Few attempts have been proposed in order to describe the statistical features and historical evolution of the export bipartite matrix countries/products. 
An important standpoint is the introduction of  a products network, namely a hierarchical forest of products that models the formation and the evolution of commodities.
In the present article, we propose a simple dynamical model where countries compete with each other to acquire the ability to produce and export new products. 
Countries will have two possibilities to expand their export: innovating, i.e. introducing new goods, namely new nodes in the product networks, or copying the productive process of others, i.e. occupying a node already present in the same network. In this way, the topology of the products network and the country-product matrix evolve simultaneously, driven by the countries push toward innovation. 
\end{abstract}

\section*{Introduction}

In the economic growth the different, endogenous and exogenous, functional requirements that let a firm  pursue products  involve the transformation and combination of tangible and intangible attributes  \cite{verona2003unbundling}, such as bureaucratic environment \cite{henisz2000institutional}, infrastructures \cite{ostrom1993institutional}, education \cite{belmonte2014italian}, etc. All these features drive either the technologic improvement in the firm production chain \cite{murtha1994country}, or the firm diversification within a country \cite{di2014diversification}, or the introduction of new products.
Current models of economic growth consider the relation between the inputs of country goods production and their effects on the overall productivity 
\cite{maddison2007world,pritchett1997divergence,aghion1998endogenous}, without taking into consideration the measure of the inputs diversity \cite{barro2003economic}.\\

\vspace{.3cm}\noindent Economic Complexity, \cite{hausmann2007structure,hidalgo2007product,hidalgo2009building,PLoSONEFFQQ,tacchella2012new,tacchella2013economic, cristelli2013measuring,zaccaria2014taxonomy, Pugliese2014}, is a new expanding field in the economic analysis, which represents a framework to measure the competitiveness of countries and the complexity of products from the national  export baskets.
The central object of study of this approach is the binary export matrix $\hat{M}$, obtained by imposing a threshold on the Revealed Comparative Advantage (RCA) \cite{balassa1977revealed} on the coutry-product trade volumes matrix.
The matrix $\hat{M}$ can be thought as the biadjacency matrix of the bipartite network \cite{Albert2002, Newman2003, caldarelli2007scale} in which one layer is represented by countries and the other by exported products.\\

In order to quantify the competitiveness of countries from the hidden information in $\hat{M}$, a new metric for countries and products has been proposed in  \cite{PLoSONEFFQQ,tacchella2012new}, overcoming flaws and problems of the seminal work \cite{hidalgo2009building}. 
The basic idea of \cite{tacchella2012new} is to define a non-linear map through an iterative  process which couples the Fitness of countries to the Complexity of products. At every step of the iteration, the Fitness $F_c$ of a given country $c$ is proportional to the sum of the exported products, weighted by their complexity parameter $Q_p$.
On the other hand the complexity $Q_p$ of a product $p$ is non linearly related to the fitness of its exporters so that products exported by low fitness countries have a low level of complexity and high complexity products are exported by high fitness countries only.\\

\vspace{.3cm}\noindent The historical evolution of $\hat{M}$ shows the development paths followed by the different countries in terms of their export flow.  It is possible to build a taxonomy network for products directly from the time evolution of the export baskets of countries \cite{hidalgo2007product,zaccaria2014taxonomy}.
 In this way the development pattern followed by different countries can be predicted as the dynamics on an evolving products network.\\

\vspace{.3cm}\noindent In this paper we present a dynamical model that describes the evolution of the export baskets of countries by implementing a minimal network model of products innovation processes, which is able to reproduce with good accuracy the main features of the observed evolution of  $\hat{M}$.
\noindent The keystone of our model is the existence of an evolving hierarchical products network in which each country occupies a subset of nodes; within this framework, the products innovation is represented by the introduction of new nodes in the products network. 
Borrowing the definition from \cite{tria2014dynamics} we distinguish between ``novelties" and ``innovation": ``innovation" is something that is new for the whole community, while ``novelty" is something known, which is new just for an individual. In this way, the novelty can be ``copied" from the near neighbours, while the process that takes to the innovation depends just on the single individual.
 
There are three main factors that drive the evolution of the export  basket of countries  and  the innovation  dynamics of products in our model: 
\begin{description}
\item[1)] the country ability to diversify its basket; 
\item[2)] the competition within a similar sector of products;
\item[3)] the ability to produce innovation with respect to the simple technological updating by adopting already developed technology by other countries. The update of the export basket of a country can take place in two ways: i) as an imitation process from other countries, introducing a novelty, ii) as the development of a brand new product, introducing an innovation. The technological updating is equivalent to the novelties introduction present in \cite{tria2014dynamics}.
\end{description}

Our model makes the products network and the $\hat{M}$ matrix evolve simultaneously,  mutually conditioning one each other: indeed, the country and product that  will evolve are chosen  on the matrix $\hat{M}$ at that time, but the kind of evolution is decided on the basis of the products network. When the country develops a new product to export, following its path on the products network, it will alter the original $\hat{M}$ matrix by modifying the products network. In this way, the efforts made by countries to develop new technologies modify in real time the path that other countries can take to diversify their own export basket.\\

The paper is organized as follows. In the section ``Methods" we first introduce the ingredients of our model, such as the data set examined (for the comparison of our model with real data) and the network of country and products; then we illustrate in details our algorithm in the subsection ``The Model". In ``Results" we analyse our results, which are going to be further commented in the section ``Discussion".

\section*{Methods}
\subsection*{Dataset}
The dataset on which we test our model is UN-NBER Sitc Rev2 \cite{webdata}, edited by \cite{FeenstraetalData}. From the import registered by the UN, the exports of the World Trade Web (WTW) is reconstructed for nearly 2577 products categories for the years interval 1963-2000. After a data cleaning procedure in order to fix some incoherences, the number of products in the analysed years interval have been fixed to 538, while the number of countries oscillates between 130 and 151, due to geopolitical changes.\\

\subsection*{The country-product network}
Economic Complexity \cite{hausmann2007structure,hidalgo2007product, hidalgo2009building, PLoSONEFFQQ,tacchella2012new,tacchella2013economic, cristelli2013measuring,zaccaria2014taxonomy, Pugliese2014} focus on the analysis of the bipartite network of countries and exported products. We start from the export volumes matrix $\hat{q}$: every entry $q_{cp}$ represents the total amount of exports in USD of the product $p$ by the country $c$. 
 In order to binarize $q$, the RCA (\emph{Revealed Comparative Advantage}, \cite{balassa1977revealed}) is calculated:
\begin{equation}\label{eq:RCA}
\text{RCA}_{cp}\definition\dfrac{\frac{q_{cp}}{\sum_{p'}q_{cp'}}}{\frac{\sum_{c'}q_{c'p}}{\sum_{c',\,p'}q_{c'p'}}}.
\end{equation}

\noindent The philosophy at the basis of Eq. (\ref{eq:RCA}) is to give a non dimensional measure of how the export basket of a specific country is organized respect to the average, comparing the impact of the product $p$ on the the export basket of $c$ respect to the impact of $p$ on the global export basket. In the light of that, we can impose a threshold on the RCA-matrix, obtaining the binary $\hat{M}-$matrix: if $m_{cp}$ is the entry for the $\hat{M}$ relative to the country $c$ and the product $p$, then

\begin{equation*}
m_{cp}=
\left\{
\begin{array}{l}
\text{if}\:\text{RCA}_{cp}\geq1\qquad 1\\
\\
\text{if}\:\text{RCA}_{cp}<1\qquad 0
\end{array}
\right.,
\end{equation*}
i.e.  only  exported products exceeding the RCA threshold appears in the basket of a country. 
An export basket in which just raw materials are over the threshold of the RCA (so that appear in the $\hat{M}$ matrix) denotes limited industrialization, while a diversified one, from highly exclusive products to most simple ones, implies a completed industrialization.
The matrix $\hat{M}$ can be thought as the biadjacency matrix of a bipartite network, in which the first layer, corresponding to the row index $c$, is composed by countries, while the second layer, corresponding to the column index $p$, is composed by the products. Links are permitted only between nodes of different layers.\\

Traditionally \cite{hausmann2010country}, the degree of the nodes, i.e. the number of links per node, are called \emph{diversification} for countries and \emph{ubiquity} for products; in terms of $\hat{M}$ they can be respectively expressed
as
\begin{equation}\label{eq:degree}
\begin{split}
k_c\definition\sum_pm_{cp};\qquad k_p\definition\sum_cm_{cp}.
\end{split}
\end{equation}

\subsection*{The model}
Our model focuses on the historical evolution of the $\hat{M}$-matrix, i.e. the biadjacency matrix relative to the bipartite undirected binary network of countries and exported products. 
The evolution of $\hat{M}$ is driven by the evolution of the products network \cite{hidalgo2007product,zaccaria2014taxonomy}, i.e. a hierarchical network based on the productivity processes such that two different products are linked if there is the possibility of passing from one to the other by a technological improvement.
The product network takes the topology of a forest in which the ``roots" represent the ancestors product, like raw materials, while the most outer leaves are the highest technology goods.\\

The tree-like topology may appear as a great simplification, in the sense that a certain production could be affected even by a ``distant" technological improvement. 
Anyway, the topology proposed has been shown, \cite{hidalgo2007product,zaccaria2014taxonomy}, to be a reliable tool able to capture the main features of countries productivity evolution; it is indeed remarkably that a so simple structure can correctly reproduce the evolution of countries diversification. 
 
A pictorial representation of the products network can be found in the left part of   the top panel of Fig. 1: the network nodes represent different products and the links are the technological relationship, while colored disks occupying a given node stand for the different countries (one color for each country) able to export the given product.

We superimposed the information contained in matrix $\hat{M}$ on the topology of the products network in order to give a more immediate interpretation of the mechanism of the export baskets evolution we are proposing.\\

Let us recall some definitions from \cite{tria2014dynamics}: in this inspiring paper focused on the nature of innovation, the authors distinguish between ``novelties" and ``innovations".
A novelty is a tool, a webpage, a song or any item you can think about that is relatively new, literally already present in the common knowledge, but not already experienced by the single agent (a person or, as in the present article, a country); in contrast, an innovation is something that has never appeared in the set of known items, so it is new for everyone.\\

In our model, a country develops its export basket progressively occupying a subset of the product network. More specifically, at each time step of the algorithm, a selected country either occupies nodes already occupied by other country or creates a new node by sprouting in the product network \emph{nearby} one of those already present in its export basket.
 Considering as the set of nodes that a country can occupy at a given time steps just the closest products its own export basket is similar to Kauffman conjecture, \cite{kauffman1995home}, about the ``adjacent possible" nature of evolution: Kauffman proposed  that the innovation process takes place only on the border  of one's own set of knowledge as items close to the borders are the most probable to be investigated and introduced.\\

Our algorithm is implemented as the sequential iteration of three fundamental substeps at each time step: the first one decides the country that will enlarge its basket; the second substep selects the product that will drive the evolution on the product network; finally, the third one will decide the path of the country basket evolution, either creating a brand new product (thus, innovating), or copying a product by adopting already developed technology by other countries (thus, introducing a novelty).\\

\begin{figure}[!h]
\centering
	\includegraphics[width=.8\textwidth]{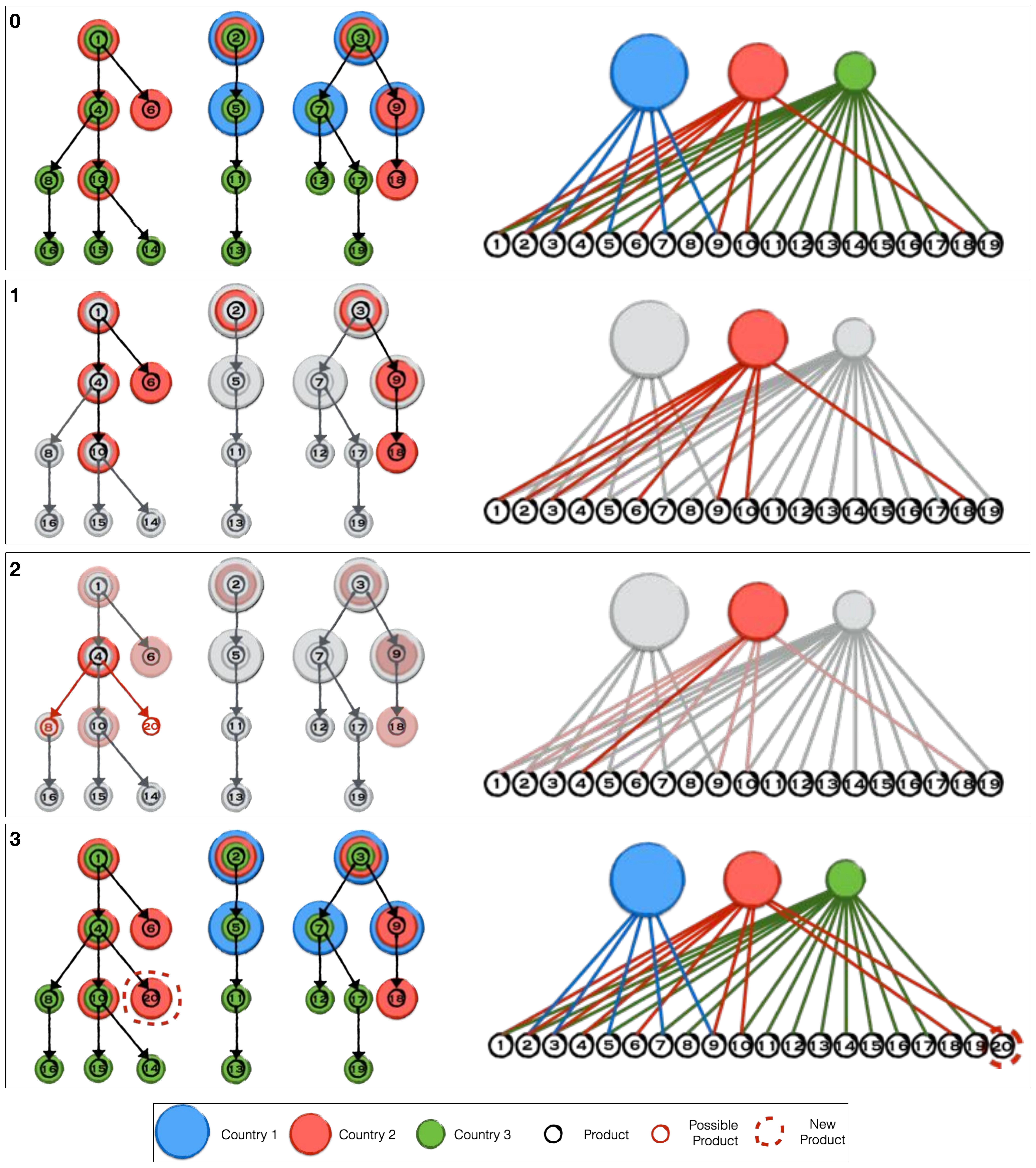}
	\caption{\textbf{Model Evolution} We propose a pictorial example of one iteration of the evolution of our model: the bipartite network of countries and exported product is on the right, while the the products network is on the left. Coloured (red, blue and green) disk represent countries, while products are circles numbered from 1 to 20; in the bipartite network  the link are coloured as the country they refer to. For completeness, we projected the information from the bipartite network on the products network, superimposing coloured disk on a given node if the country is able to export the single product. In the top panel of Fig. \ref{fig:modelevolution} we present the initial condition, in which the red, blue and green countries have their export products and occupy different nodes in the products network. In the second panel of Fig. \ref{fig:modelevolution}  the first substep of the algorithm is taken: following the recipe described in the section ``The method" the red country is selected. The products in ``red" export basket are highlighted both on the bipartite network (on the right) and on the products network (on the left).
At the second substep, among the products in the ``red" export basket, ``4" is selected in the third panel of Fig. \ref{fig:modelevolution}. Together with the previous step, we have selected a link  in the bipartite network. On the left, the possible choices on the products network for the third substep: since ``10" is already in the ``red" export basket, it cannot be selected as the final target for the evolution, so the only possibility left are product ``8" (already produced by ``green") or ``20", a brand new product.
In the bottom panel of Fig. \ref{fig:modelevolution} the thirds choice has been taken and ``20" is a new product in the products network.
}\label{fig:modelevolution}
\end{figure}

In details, the 3 sequential substeps of our model are the following:

\begin{description}
\item[1) The country selection: Divesification.]

At the first substep, a country $c$ has a probability to be selected

\begin{equation}\label{step1}
P_1(c)\sim\,k_c^\alpha
\end{equation}
where $k_c$ is its diversification, as defined in Eq.(\ref{eq:degree}), and $\alpha>0$ is a parameter of the model. 
Normalization of the all countries is imposed to Eq.\eqref{step1} in order to evolve a single country at each time step.
This first substep is similar to the generalization of the preferential attachment 
 presented in  \cite{Krapivsky2001}, with the difference that here we select a node on one layer of a bipartite network, while in \cite{Krapivsky2001} a node was selected in a monopartite one.
 
Equation \eqref{step1} says that countries with a diversified export basket have a higher probability to be chosen:
it implements the idea that the diversification can be taken as a good proxy for efforts a country makes in order to evolve its export basket (a wider discussion is developed in \cite{hesse2006export}). 
The selected country is the one performing the evolution of matrix $\hat{M}$ in the next two substeps.

In the second panel of Fig. \ref{fig:modelevolution} the first step is pictorial shown: among the three countries (red, blue and green), represented by different disks 
in the upper layer of the bipartite network, the red one is selected and its export basket in the product network highlighted. Note that the diversification $k_c$ is the number of nodes occupied by each country, so we have $k_{\text{red}}=8$, $k_{blue}=5$ and $k_{\text{green}}=16$.
\item[2) The evolving product selection: Competition.]
Once the country $c$ has been chosen, we have to select a product already in its export basket, from which either moving towards an unoccupied existing neighbouring nodes or sprouting a brand new one.
We select such a  product $p$ with a probability:
\begin{equation}\label{step2}
P_2(p|c)\sim k_p^\beta
\end{equation}

where $k_p$ is the ubiquity, as defined in Eq.(\ref{eq:degree}) and $\beta>0$ is the second parameter of the model.
Similarly to Eq.\eqref{step1},  Eq.\eqref{step2} implements the generalization of the preferential attachment criterion of \cite{Krapivsky2001}, applied to the product layer of our bipartite network.\\

The idea is that the more producers, the harder the efforts on renovation, the more possibilities of improving the export basket may come from the most ubiquitous products. 
In effect, similar behaviours has been shown by experimental evidences, \cite{clauset2009power}. \\ 

The third panel of Fig. \ref{fig:modelevolution} illustrates such second substep: from the commodities present in the ``red" country export basket, $p=``4"$ is selected (while on the bipartite network $k_p$ is the number of countries linked to $p$, in the  products network it is represented by the number of differently coloured disk occupying the selected node).

\item[3) Target product selection: Innovation against Novelty.]
In the previous substeps, we have selected the country $c$ and the product $p$ (in the pictorial representation of Fig. \ref{fig:modelevolution} the ``red" country and the product ``4") performing the last evolution substep. 
This choice was based on the properties of the matrix $\hat{M}$ at that time steps; more precisely on diversification and ubiquity defined in Eq. \eqref{eq:degree}.
So far, no information from the products network topology has been used.\\

Let us now consider the position the chosen $p$ occupies in the products network.
There are two options: either introducing a new node in the products network, i.e. innovating, or evolving along the links already present in the products network by introducing in the export basket a product neighbour of $p$ already exported by other countries, i.e. introducing a novelty.

The probability of copying novelties from others will increase with the number of countries that already export the given good.
In this way we implement the idea that it is much easier to acquire close technology. 
On the other hand, the innovation process of proposing brand new products will need an extra parameter.\\

As possible novelties we consider all the first neighbours $p'$ of $p$ in the product network which are not already present in the basket of $c$.
At the same time let us call $p^*$ a possible brand new product sprouting out of $p$. 
We call  $\tilde{p}$ the generic element  of the set obtained by the union of $p'$ and $p^*$.
In this third substep  we  select  a  single element $\tilde{p}$ of this set with the probability given by:
\begin{equation}\label{step3}
P_3(\tilde{p}|c,p)\sim(k_{\tilde{p}}+k^0)^\gamma,
\end{equation}
where $\gamma,\,k^0>0$ are the last two parameters of the model and $k_{\tilde{p}}$ is the ubiquity of $\tilde{p}$; clearly $k_{p^*}=0$.
The quantity $k^0>0$ is a necessary an offset permitting even the artificial product $p^*$ to be selected. The $k_{\tilde{p}}$ term, makes the probability of \coniglietti{copying} other ``accessible" products made already by other country larger than introducing an innovation.\\

The third panel of Fig. \ref{fig:modelevolution} illustrates this third substep: in the second panel of the same Fig.\ref{fig:modelevolution} we selected the product $p=``4"$.

Now the possibilities are between either selecting the  product $``8"$, already produced by the ``green" country, or introducing a brand new product $p^*=``20"$. In the picture we represent pictorially this last event.

\end{description}

In our algorithm we iterate these three substeps until the number of products in the networks is the same of the observed matrix we want to reproduce, i.e. $538$ for the examined data.
Summarizing the parameters of the model are $\alpha,\,\beta,\,\gamma,\,k^0$.

\subsubsection*{The density saturation}

\begin{figure}[!h]
\begin{center}
\includegraphics[width=1.\textwidth]{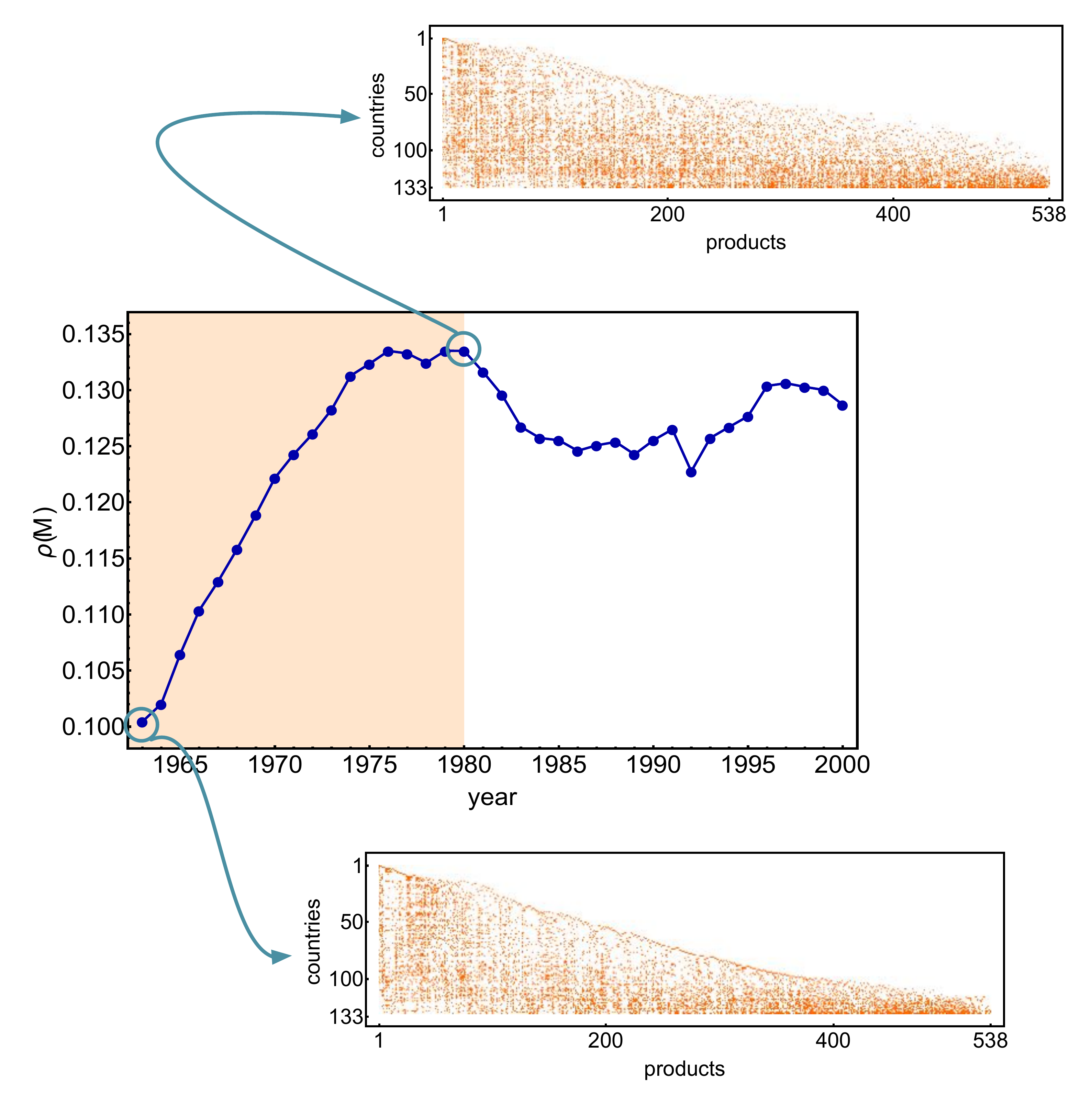}
\end{center}
\caption{\small The density evolution for the dataset \cite{webdata, FeenstraetalData}. It is possible to observe the density increasing until a certain value $\rho_\text{cr}\sim0.13$ for the year $\sim1975$; our model follows a similar behaviour, limiting the evolution to the export of existing products once the number of nodes in the products network reaches the number of observed products in the real network.}
\label{fig:Rhoev}
\end{figure}

For all the simulated values of the parameters once the number of products introduced in the network reaches the same number of the observed bipartite network, the density of links is however smaller than the observed one.
Consequently starting from this time-step we will set $k^0=0$ in order to prevent the creation of new products and permit only the introduction of novelties increasing in this way the density of bipartite network.
We stop the iteration of the algorithm when the density of the country-product bipartite network saturates to the observed one is $\rho_{cr}\simeq0.13$.
Interestingly we observed that a similar behaviour is shown by real data as illustrated by Fig. \ref{fig:Rhoev}.
In this figure we can appreciate that the density of the bipartite network increases from year to year up to $1975$ and then saturate to an approximate constant values.
One possible explanation of this phenomenon is probably the fact that products categories of Sitc Rev.2 have been formalized in 1980 and data for the years before 1980 have been converted to the Sitc Rev. 2 by the Sitc Rev.1 and some data may have been lost in merging the datasets.

\subsubsection*{Initial Conditions}
In order to render the final results of our model less sensible to the initial conditions, we compose the initial exports basket of each country as a Bernoulli trial.
We start with a given number $N_\text{roots}$ of initial products, hereby called {\em roots}.
We assign each product of this set to each country with a constant probability $P_0$, independently of the other countries. 
Consequently each country will have a random number of initial products selected from the set of roots, whose probability distribution is binomial with mean value $N_\text{roots}P_0$ and standard deviation $\sqrt{N_\text{roots}P_0(1-P_0)}$.
At this level the product network is a set of disconnected nodes, represented  by the roots; at the first step of the evolution dynamics the product network will evolve from this initial condition as a branching process in which the branching event is represented by innovation.

\subsubsection*{Exploring parameter space}

As explained in the Supplementary Material, in order to find the best range for the values of the parameters characterizing both the initial condition, $(N_\text{roots}, P_0)$, and the evolutionary model, $(\alpha,\beta,\gamma, k^0)$,  we tested for each 
choice of the first pair, in the range of reasonable values of $N_\text{roots}$ from $10$ to $40$ and of $P_0$ from $0.2$ to $0.6$, a wide range of the dynamical parameters. We found that the performance of the algorithm weakly depends on the precise choice of $(\alpha,\beta,\gamma, k^0)$ around reasonable values for the exponents $\alpha,\beta,\gamma$ around the unitary value, and for $k^0$ around the minimal observed value $k^0=4$ for the product ubiquity $k_p$ in real data.
The slightly best performing set of values is however found to be $N_\text{roots}=20$, $P_0=0.3$, $\alpha=1.55$, $\beta=0.8$, $\gamma=0.3$, $k^0=4$.

\section*{Main Results}

We compare the matrices $\hat{M}$ generated by our algorithm for different values of the parameters ($\alpha,\,\beta,\,\gamma,\,k^0$) with the observed one using several non-trivial quantities characterizing the features of binary bipartite networks.
In particular we will use the countries Fitnesses and products Complexities distributions, as defined in  \cite{PLoSONEFFQQ,tacchella2012new}; nestedness \cite{AlmeidaNetoetalNest}; assortativity \cite{NewmanAssortativity}; motifs for bipartite networks \cite{checkerboardsscores,saraccoetal}.

\begin{figure}[!h]
\begin{center}
\includegraphics[width=1.\textwidth]{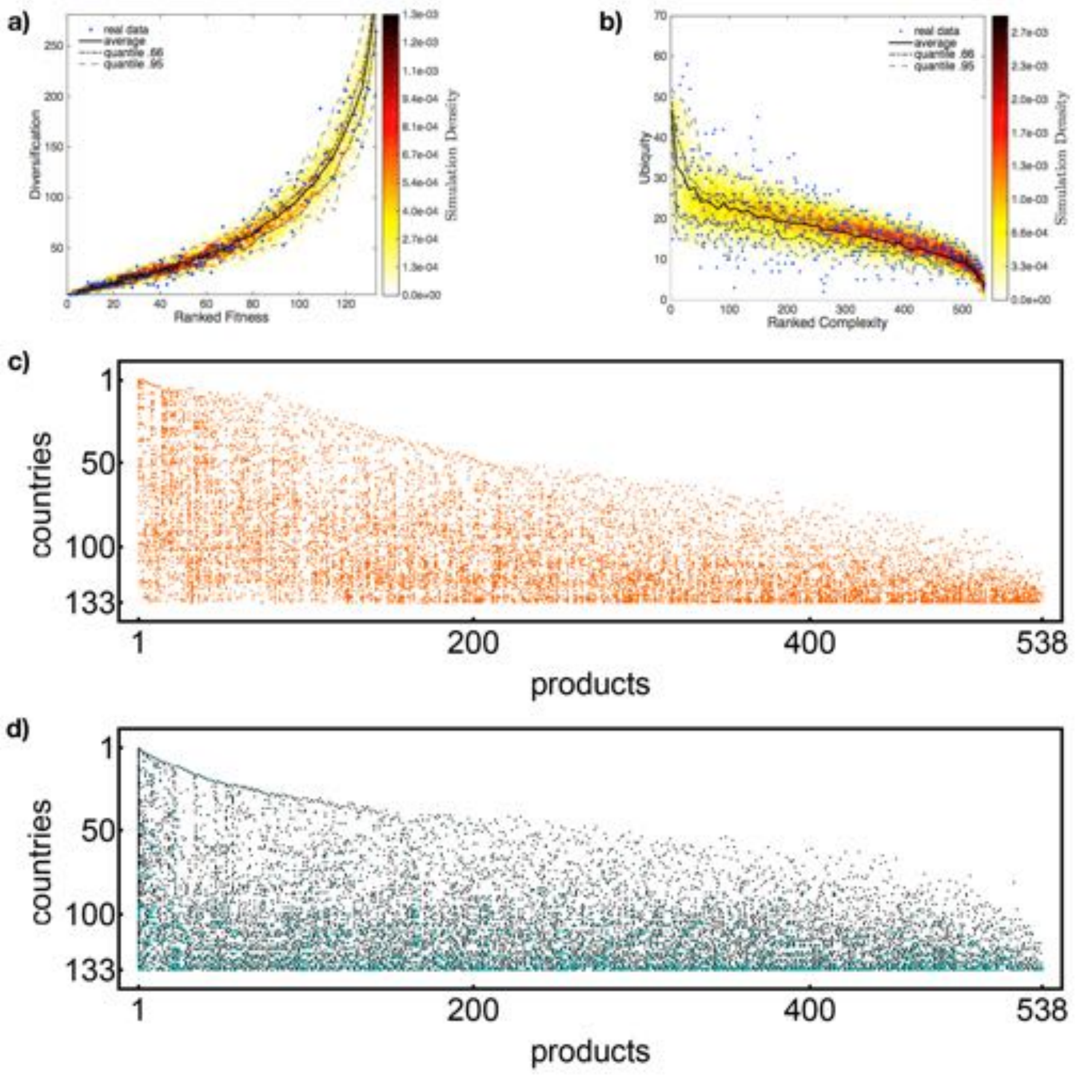}
\end{center}
\caption{\small In Fig. \ref{fig:mah}(a) the scatter plot of Fitness ranking against countries diversification, while in Fig. \ref{fig:mah}(b) the one for  Quality ranking against products ubiquity; the blue points represent the observed values (for the year 1980 from the dataset of \cite{webdata, FeenstraetalData}). The black line represents the average value on the simulations, while the grey lines bind the area between the second and the first 3-quantiles (dot-dashed) and between the 975th and 25th permilles (dashed). The data obtained are for initial conditions $N_\text{roots}=20$ and $P_0=0.3$  and parameters $\alpha=1.55,\,\beta=0.8,\,\gamma=0.3,\,k^0=4$. In the $\sim82\%$ the observed data fall into the area between 975th and 25th permilles for the fitness distribution,  $\sim75\%$ for the quality distribution. In Fig. \ref{fig:mah}(c) the original matrix for 1980 from the dataset of \cite{webdata, FeenstraetalData}; in Fig. \ref{fig:mah}(d) one of the synthetic matrix for initial conditions $N_\text{roots}=20$ and $P_0=0.3$  and parameters $\alpha=1.65,\,\beta=1.1,\,\gamma=0.6,\,k^0=4$.}
\label{fig:mah}
\end{figure}

\begin{figure}[!h]
	\centering
	\includegraphics[width=\textwidth]{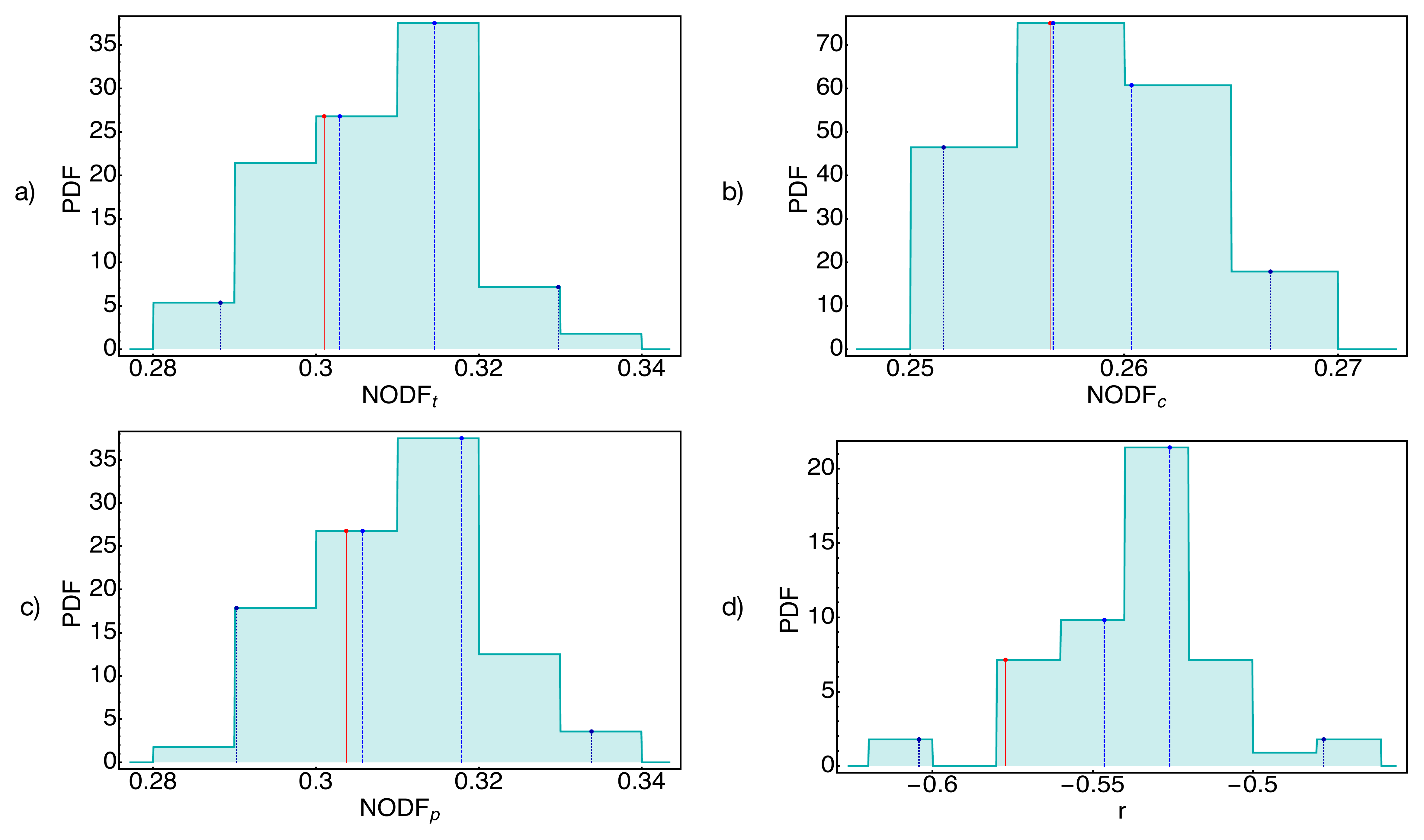}
	\caption{\small The distributions for the nestedness values (obtained employing NODF, the definition by \cite{AlmeidaNetoetalNest}) and assortativity index $r$ (obtained employing the definition by \cite{NewmanAssortativity}) for  50 simulations with initial conditions $N_\text{roots}=20$ and $P_0=0.3$  and parameters $\alpha=1.55,\,\beta=0.8,\,\gamma=0.3,\,k^0=4$. In Fig. \ref{fig:nodfandr}(a) the total NODF, in Fig. \ref{fig:nodfandr}(b) the NODF for rows and in Fig. \ref{fig:nodfandr}(c) the one for columns. The red line is the observed value for the year 1980 from the dataset of \cite{webdata, FeenstraetalData}, the blue dashed lines bind the area between the second and the first 3-quantiles, while the dark blue dots mark the area between between the 975th and 25th permilles. For the 4 distributions, real values easily fit in the 95\%; anyway, for NODF values the real values lie just outside the central third of the probability. Notice the similar distributions for $\text{NODF}_t$ and $\text{NODF}_p$, as explained in Eq. \reff{eq:NODFapprox}. In Fig. \ref{fig:nodfandr}(d) the distribution for the assortativity values (obtained employing the definition by \cite{NewmanAssortativity}): Even if the distribution is quite weird, the value measured on the real matrix is just outside the area containing the 33\% of the distribution.}
	\label{fig:nodfandr}
\end{figure}

\begin{figure}[h!]
	\centering
		\includegraphics[width=.7\textwidth]{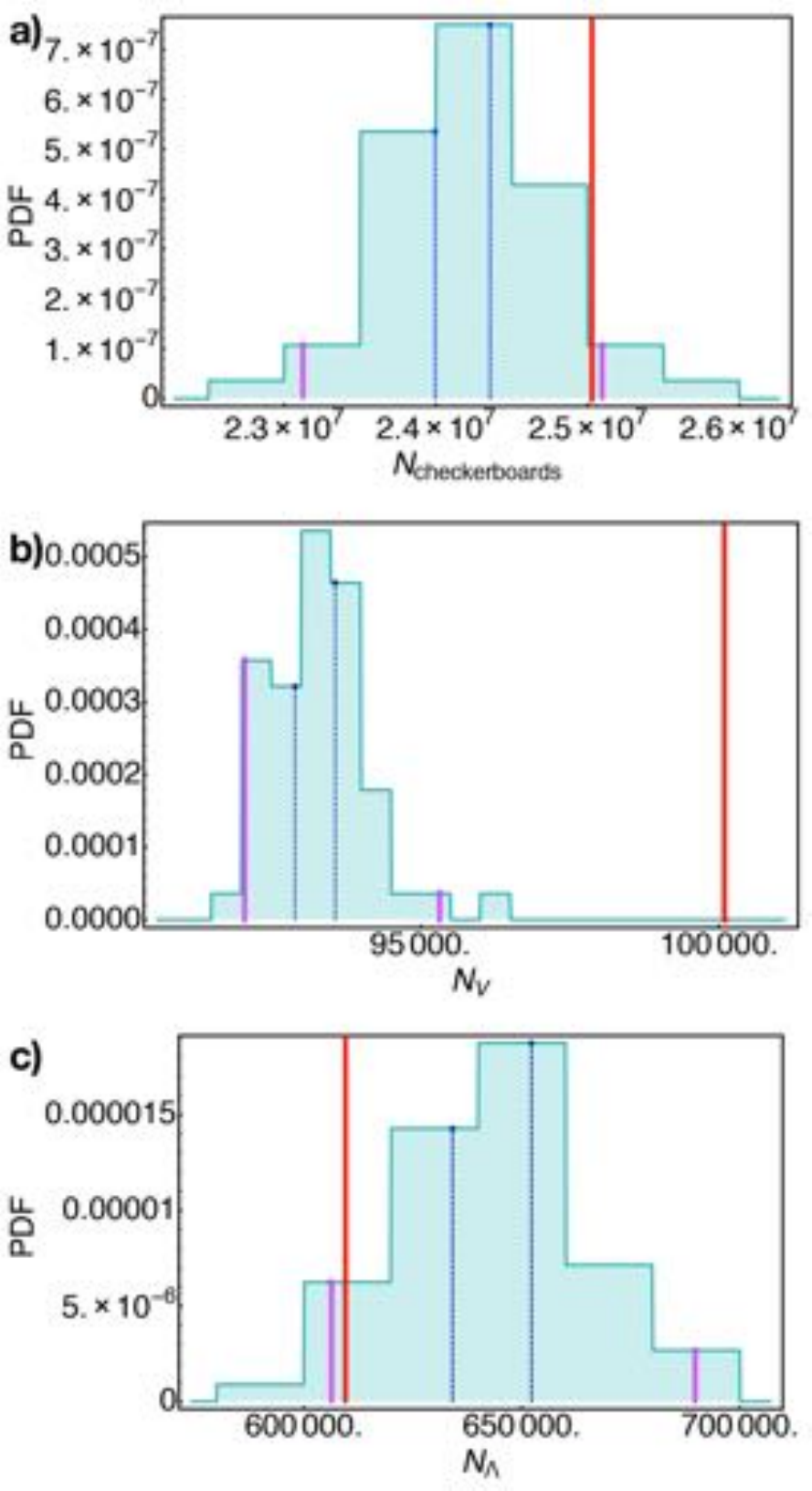}
	\caption{\small The distribution for the number of checkerboards(Fig. \ref{fig:motifs}(a)), $V-$ (Fig. \ref{fig:motifs}(b)) and $\Lambda-$motifs (Fig. \ref{fig:motifs}(c)), obtained from the simulation with initial conditions $N_\text{roots}=20$ and $P_0=0.3$ and parameters $\alpha=1.55,\,\beta=0.8,\,\gamma=0.3,\,k^0=4$. The red line is the observed value for the year 1980 from the dataset of \cite{webdata, FeenstraetalData}, the blue dashed lines bind the area between the second and the first 3-quantiles, while the dark blue dots mark the area between between the 975th and 25th permilles. While \ref{fig:motifs}(a) and \ref{fig:motifs}(c) shows that the number of checkerboards $\Lambda-$motifs are reproduced by the model, in \ref{fig:motifs}(b) the real value lies outside the 95\% of probability; the presence of a hierarchy in the set of products captures the right values of checkerboards and $\Lambda-$motifs, but it is not enough to reproduce the $V-$motifs.}
	\label{fig:motifs}
\end{figure}

\begin{description}
\item{\textbf{Fitness and Complexity}}  In spite of the simplicity of our the model, there is a remarkably good agreement between simulations and real data for Fitness/Complexity (for details about the definition of Fitness and Complexity, see the Appendix).
In particular the shape of the scatter plot for countries fitnesses (products complexities) ranking against countries diversifications (products ubiquities) reproduces well the behavior in the real data; the result is shown in Fig. \ref{fig:mah}(a) (Fig. \ref{fig:mah}(b)). 
It is possible to see that our algorithm is able to reproduce the ``shape" of the original matrix data (blue dots) within the 95\%, which is a remarkable result, since these peculiar trends are derived by the highly non-linear algorithm for Fitnesses and Complexities.

\item{\textbf{Nestedness}}  The concept of nestedness relates to how much a row (or a column) is subset of the others.
 In this sense, it is a way to evaluate the ``triangularity" of the binary matrix $\hat{M}$.
Among the different definition of nestedness \cite{AlmeidaNetoetalNest, BastollaetalNest, MaritanetalNest, MunozNest}, we opted for the NODF (\emph{Nestedness measure based on Overlap and Decreasing Fill}), presented in \cite{AlmeidaNetoetalNest}.
This choice is motivated by the fact that is independent form the order of the elements and it is particularly intuitive.
The final value of nestedness is the sum of the contributions by the columns (i.e. products) and the rows (i.e. countries), weighted by the possible couple of elements (for details about the definition of the NODF, see the Appendix); 
for completeness, we analysed the contributions from countries and products separately, i.e. $\text{NODF}_c$ and $\text{NODF}_p$ respectively, as well as the total contribution, $\text{NODF}_t$.\\ 

As it can be seen in the Fig. \ref{fig:nodfandr}, panels c, the $\text{NODF}_p$ of real data is well replicated by our algorithm.
This means that our model is able to catch the main features of the hierarchical organization of products; this result is due to the assumption of the presence of a products network, i.e. a hierarchical structure among products. 
The result for $\text{NODF}_c$, in the panel b of Fig. \ref{fig:nodfandr}, is much more non trivial: in this case the values from the simulations reproduce the same quantity for the real matrix, even though no explicit structure was imposed on the set of countries. \\

As a matter of fact countries, by following productivity paths on the products network, impose a nested structure to countries as a consequence of the i) hierarchical structures of  products, and ii) the mechanism of the products network evolution .
Since poorly diversified countries cover a subset of the products network of the most diversified  one, the  matrix $\hat{M}$ appears nested respect to rows too.

\item{\textbf{Assortativity}} As proposed in \cite{NewmanAssortativity}, assortatitvity is a measure of how much nodes link nodes with similar degree (more details can be found in the Appendix). Since the poorly diversified countries focus their export toward the most ubiquitous products, the export  bipartite network is disassortative, i.e. a network in which low degree nodes links to high degree ones. Our model provides a value inside the 95\% of probability, but, respect to other measurements, the assortativity shows a worse agreement between the simulations and real data, as can be seen in Fig.  \ref{fig:nodfandr}(d). Actually, this behaviour is probably due to a sort of  ``second order" effect: the most diversified countries do exports even the most ubiquitous products, but their export basket is nevertheless biased towards the highest quality products. In effect, the distribution in Fig. \ref{fig:nodfandr}(d) shows that the model provides a less disassortative bipartite network respect to the real one (in red in the figure).
 
\item{\textbf{Motifs}}. Let us represent an entry in the $M-$matrix as a square which is empty ($\square$) if $m_{cp}=0$, while it is filled ($\blacksquare$) for $m_{cp}=1$. The checkerboard score \cite{checkerboardsscores}, i.e. the number of pattern $\checker$ and $\antichecker$ inside the $M-$matrix counts the mutually exclusive exported products for two different countries. As shown in Fig. \ref{fig:motifs}(a), the total number of checkerboards is finely reproduced by the model; the agreement with real data means that the evolutionary algorithm well replicates the diversity of technological development roads followed by countries.\\
In \cite{saraccoetal,saraccoetal2} several motifs for bipartite networks have been proposed in order to uncover the structural properties of the system at hand; in the following we will consider just the simplest ones, $V-$motifs and $\Lambda-$motifs. In few words, they represent the number of co-occurrence of products in the basket of two different countries (the $V-$motifs) or the co-occurrence of countries in the set of the producers of two different products ($\Lambda-$motifs); more details on the definitions of the motifs can be found in the Appendix.\\
In  most of the cases the real value of the number of $\Lambda-$motifs falls at the borders of the first 3-quantile and far inside the area contained between the 25th and the 975th permilles (that is, the area containing the 95\% of the probability around the mean value), see Fig. \ref{fig:motifs}(c).\\ 
On the countries set we did not assume any kind of structure and it is probably the cause for not reproducing the total number of $V-$motifs, see Fig. \ref{fig:motifs}(b). In effect, simulated data always fall out of the 95\% of the probability around the median of real data: apparently the evolutionary paths described by the products network is not enough for reproducing $V-$motifs.

\end{description}

The first 3 measurements, i.e (i) the fitness and complexity scattered against the nodes degrees, (ii) the nestedness and (iii) the assortativity, explicit the hidden information encoded in the triangular shape of the matrix $\hat{M}$, while the motifs carry part of the topological information of the bipartite network. Our evolutionary model is able to replicate those measures, showing that the products network created by our algorithm  can be a good starting point to better understand the hidden forces which produce the main characteristics of the export bipartite network structure.

\subsection*{The evolution}
\begin{figure}[h!] 
	\centering
		\includegraphics[width=.7\textwidth]{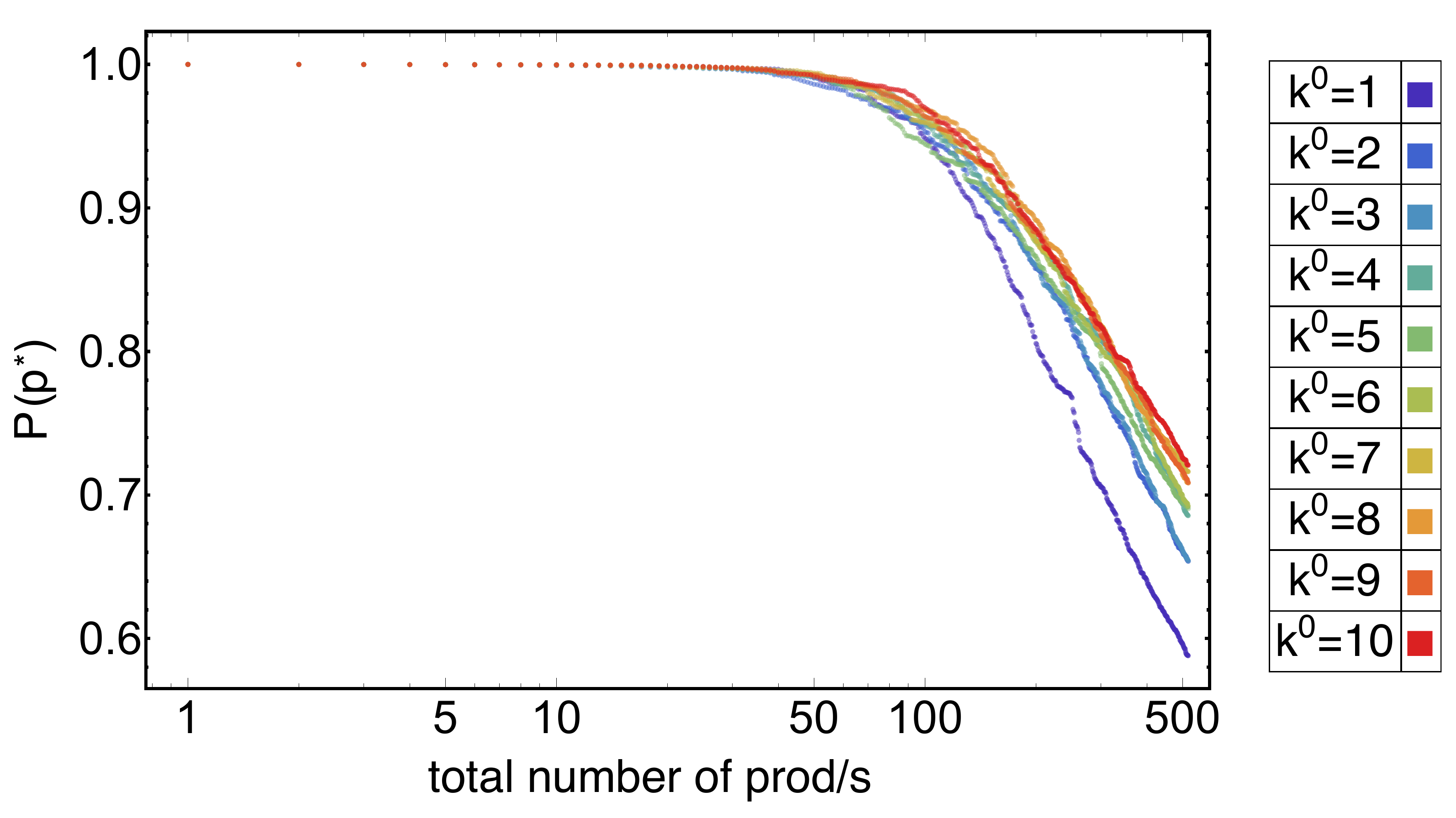}
	\caption{\small The evolution of the probability of innovation depending on $k^0$ (on the horizontal axis the total number of products at the time). It is possible to observe two different phases: a first period of discoveries, when the probability of innovating is close to 1 and a second period in which the spread of the novelties increase. It is worth noticing that all slopes for $k^0>1$ cluster together.}
	\label{fig:innovation}
\end{figure}

The total probability of introducing an innovation can be obtained assembling Eqs. \reff{step1}, \reff{step2} and \reff{step3}: 
\beq
\begin{split}
P(p^*)=&\sum_{c,p}P_1(c)P_2(c|p)P_3(p^*|c,p)\\
=&\sum_c\dfrac{k_c^\gamma}{\sum_{c'}k_{c'}^\alpha}\cdot\sum_p\dfrac{k_p^\beta}{\sum_{p'}m_{cp'}k_{p'}^\beta}\cdot\dfrac{(k^0)^\gamma}{(k^0)^\gamma+\sum_{p''}A_{pp''}(k^0+k_{p''})^\gamma}
\end{split}
\eneq
The evolution of the probability of innovating respect to the total number of products exported at that time has been reported in Fig. \ref{fig:innovation}; different colors represent different values of $k^0$, while the other parameters $(\alpha, \beta, \gamma)$ have been fixed respectively to (1., 1.6, 0.4). 
At the early stage of the evolution dynamics, the probability of innovating is obviously close to 1, as the number of branching is negligible with respect to the number of roots. This happens until the total number of products is around $\sim50$ for all the value examined of $k^0$. After that threshold, the possibility of following path already developed by others countries reaches a higher value.\\ 
The plot in Fig. \ref{fig:innovation} shows that there are two phases: a first period of ``great discoveries" (until the total number of products is $\leq50$) in which the topology of the early products network is shaped, and a second period in which the technology innovations diffuse, under the form of novelties, among countries. Note also that slopes for different values of $k^0$ cluster: while $k^0=1$ is almost alone, sketching a steep trajectory, other values are next to each other.\\ 

\begin{figure}[h!] 
	\centering
		\includegraphics[width=.7\textwidth]{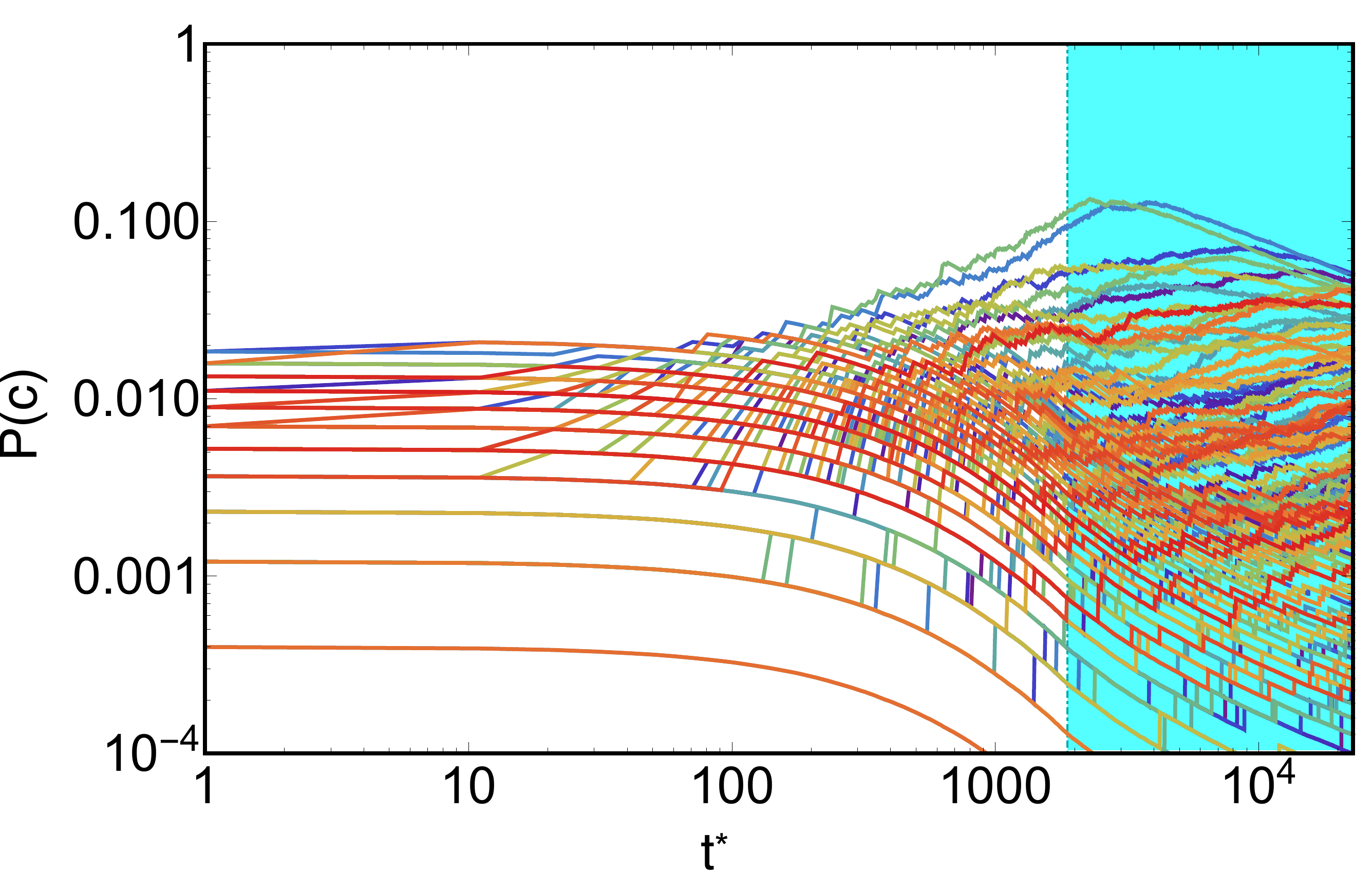}
	\caption{\small The evolution of the probability of selecting every country; on the horizontal axis there is the simulation time. Until the saturation regime (the cyan area) few countries start increasing their probabilities of being selected with the increasing of their diversification, to the detriment of the poor diversified countries, whose probabilities are pushed lower. In the saturation period, the mid-diversified countries enlarge their export basket, boosting their probabilities of being selected, while highest diversified countries are restrained; in this way the gap among countries reduces.}
	\label{fig:probc}
\end{figure}

A similar, but different, discussion can be made about the evolution of the probability of countries selection, given by Eq.\reff{step1}. In Fig. \ref{fig:probc} the evolution of this probability for single country is plotted (differently from Fig. \ref{fig:innovation} on the horizontal axis the evolution time is plotted). This plot clearly shows the effect of the late saturation time interval (when, as explained above, $k^0=0$) on the selection probability of a single country: from an initial almost uniform diversification, few countries start becoming more and more diversified, increasing in this way their probability of being selected at further times, and, consequently, reducing the same probability of  poor ones due the normalization over all countries. In the saturation time interval, represented by the cyan area in Fig. \ref{fig:probc}, we note a shrinking of the selection probability distribution over countries, due to the prevention of further innovation. Since just novelties are permitted at this final stage mid-diversified countries improve their chances of diversifying, while already developed are restrained by the fact that they are already extremely diversified and novelties occur with less frequency.\\

\begin{figure}[!h] 
	\centering
		\includegraphics[width=.7\textwidth]{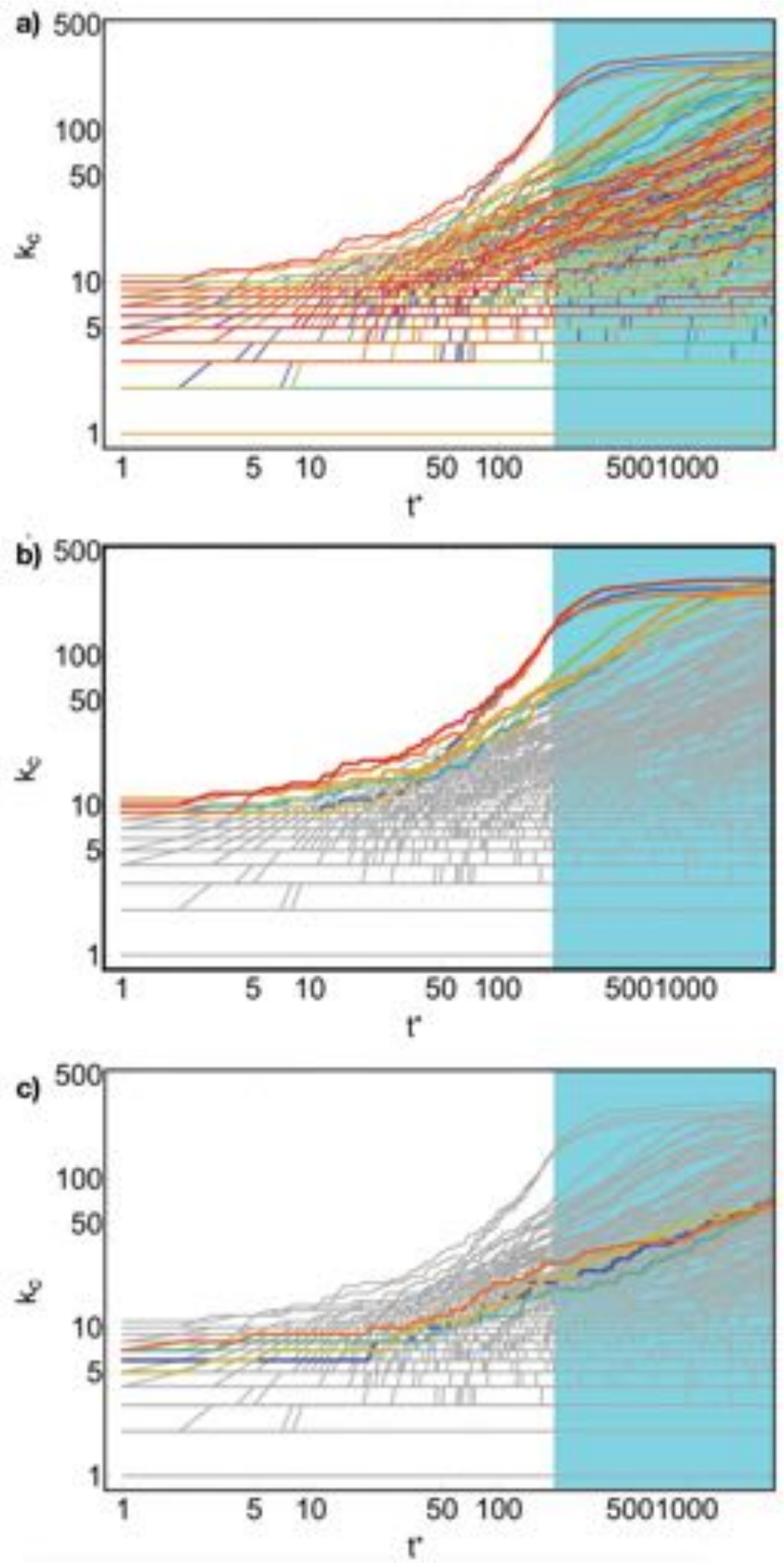}
	\caption{\small The evolution of the diversification against the simulation time. In Fig. \ref{fig:diversification}(a) all countries, but those whose initial conditions have been particularly unlucky, evolve, boosting their diversification; in the cyan area, i.e. during the saturation regime, highly diversified countries experience an evolution different from others. In effect, it is possible to observe, once focusing on the highest fitness countries, Fig. \ref{fig:diversification}(b), that all of them experience a growth with a $S-$curve profile, which is peculiar of biologic and economics system with finite resources. Note the other less diversified countries, Fig. \ref{fig:diversification}(c), this phenomenon is not present; effectively the main target of the saturation regime is reducing the gap between the diversification of fully developed and developing countries.}
	\label{fig:diversification}
\end{figure}

The same effect can be observed directly on the distribution of diversification over countries; Fig. \ref{fig:diversification} shows that there is an abrupt raise in the diversification during the evolution for few countries, while others experience a slower evolution (again, on the horizontal axis there are the time steps of the simulation). Moreover, the most diversified countries, i.e. those which feel strongly the decrease in the probability of being selected due to the saturation, show an $S-$shaped profile of the evolution curve, Fig. \ref{fig:diversification}(b): after a steep raising slope, the saturation time determines a slow increase towards a limiting value. Figure  \ref{fig:diversification}(c) illustrates that not all countries show such a $S-$shaped behaviour due to the presence of the saturation stage, but just those with high diversification; others do not occupy the products network enough for feeling the difference between the products network growth and the saturation regime.
The presence of such an $S-$shape curve is also typical in evolutionary models in biology and socio-economics, when resources are limited, \cite{Rogersdiffusion, barbosa2009}.

\section*{Conclusion}
The main target of the Economic Complexity approach, \cite{hausmann2007structure,hidalgo2007product, hidalgo2009building, PLoSONEFFQQ,tacchella2012new,tacchella2013economic, cristelli2013measuring, zaccaria2014taxonomy, Pugliese2014}, is to unveil, through the information contained in the binary bipartite network of countries and exported products, the productive capabilities of different countries and the industrial hierarchical space. Quite surprisingly with respect to some celebrated economic theories, the biadjacency matrix $\hat{M}$, defining the bipartite network, exhibits a peculiar approximately triangular shape. This shows that the most diversified countries, i.e. those able to export a wide class of different products, are the only ones able to export both the most technologically advanced goods as well as the simplest ones. On the contrary poorly diversified countries usually export only ubiquitous products, which in general bring a low level of industrial complexity. This new approach permitted to reach many interesting results about the competitiveness of countries and the complexity of products. In particular the construction of a products network, defining a hierarchy among products, permits to determine the different paths followed by countries in the product space to develop their export basket \cite{hidalgo2007product,zaccaria2014taxonomy}.\\

In this framework we proposed a simple dynamic evolution algorithm that, starting from general  initial conditions
is able to reproduce the main features of the observed countries/products bipartite network, as different measures for the ``triangular" shape of the observed $M-$matrix, in a wide range of the parameters around reasonable values. The central ingredient of the evolutionary model is the progressive and self consistent construction of the product network, encoding the different steps of the technological progress. \\
Our model provides the simultaneous evolution of the matrix $\hat{M}$ and the products network; the dynamical evolution of the latter at the same time drives and is driven by the evolution of the matrix $\hat{M}$, as the technological evolution depends tightly on the productive capabilities of the different countries, i.e. on the nodes each country occupies in the products network.\\

The proposed model is able to reproduce the main features of the observed bipartite network for a wide range of parameters. In particular we compared the ``shape" of the matrix $\hat{M}$, as encoded in the Fitness/Complexity algorithm\cite{PLoSONEFFQQ,tacchella2012new, tacchella2013economic,cristelli2013measuring,Pugliese2014}; the measures of the nestedness (in the definition proposed \cite{AlmeidaNetoetalNest}) and the assortativity, \cite{NewmanAssortativity}; some motifs, like the checkerboards patterns, \cite{checkerboardsscores} and the $V$ and $\Lambda$, \cite{saraccoetal,saraccoetal2}.\\
For the range of parameters examined we find the observed values of all quantities with a single minor exception ($V-$motifs) inside the 95\% of the distribution of simulated data.\\

Our model is meant to be a first step in the direction of a dynamical network approach to the processes of countries innovation and competition on the exports. There are several possible directions of improvement, as implementing different evolutionary rules for the construction of the product network and the modeling of the countries dynamics on it. 
 
Another possible direction could be the introduction of an appropriate random process of losing products from the export basket, simulating exogenous phenomenon as the progressive obsolescence of ``old" products or the presence of socio-political factors (as wars, traditions, political resolutions, etc.). All these (and other) possible approaches are going to be studied in following works.

\section*{Acknowledgements}
This work has been partly funded and supported by Italian PNR project ``CRISIS-Lab'' and EU Project nr. $611272$ ``GROWTHCOM''.
We acknowledge Tiziano Squartini, Andrea Tacchella  for many interesting and important discussions.
We are also grateful to Claudia Testi for useful cross-checks.

\section*{Author contributions}
R.D.C. \& F.S. developed the model; F.S. \& R.D.C. carried out the computer simulation and made the figures. F.S., R.D.C. \& A.G. wrote the paper \& L.P. revised the main manuscript text.

\section*{Additional Information}
Competing financial interests: The authors declare no competing financial interests.

\bibliographystyle{naturemag}

\newpage

\section*{Appendix}\label{sec:Appendix}

\subsection*{Fitness and Complexity}

In order to extract the information contained in the $\hat{M}$-matrix, the authors of \cite{PLoSONEFFQQ,tacchella2012new, tacchella2013economic,cristelli2013measuring}, overcoming flaws present in the seminal works \cite{hidalgo2007product, hidalgo2009building}, propose a metric for countries and products, the celebrated Fitness and Complexity algorithm: this recursive and non linear algorithm is a sort of PageRank applied to bipartite networks, where Fitness is the quantity for countries, while Complexity is the one for products. The idea at the the basis of the algorithm is that highest fitness countries are those which are able to export the highest number of the most exclusive products, i.e. those with the highest complexity. In particular, the Fitness $F_c$ for the generic country $c$ and Quality $Q_p$ for the generic product $p$ at the $n-$th step of iteration, are defined as
\begin{equation}\label{eq:FFQQ}
\left\{
\begin{array}{c}
\tilde{F}^{(n)}_c=\sum_p m_{cp} Q^{(n-1)}_p\\
\\
\tilde{Q}^{(n)}_p=\dfrac{1}{\sum_c m_{cp} \frac{1}{F^{(n-1)}_c}}
\end{array}
\right.
\rightarrow
\left\{
\begin{array}{c}
F^{(n)}_c=\dfrac{\tilde{F}^{(n)}_c}{\langle \tilde{F}^{(n)}_c\rangle}\\
\\
Q^{(n)}_p=\dfrac{\tilde{Q}^{(n)}_p}{\langle \tilde{Q}^{(n)}_p\rangle}
\end{array}
\right.,
\end{equation}
where the symbols $\langle\cdot\rangle$ indicate the average taken over the proper set. The initial condition are taken as $F_c^0=Q_p^0=1\,\,\forall c\in N_c,\,\forall p\in N_p$, where $N_c$ and $N_p$ are the number respectively of countries and products 
(the convergence of the algorithm described by Eqs.(\ref{eq:FFQQ}) depends on the shape of the matrix $\hat{M}$, as it has been discussed in \cite{Pugliese2014}). 
\subsection*{Non trivial benchmarks}

\subsubsection*{Fitness and Quality distributions}
Using Fitness and Complexity it is possible to reveal several non-trivial properties of the $\hat{M}$-matrix: the very first observation is that, once reordered rows and columns respectively by fitness and complexity, the $\hat{M}$-matrix shows a peculiar triangular form, already observed in biological systems, \cite{AlmeidaNetoetalNest, BastollaetalNest, MaritanetalNest, MunozNest}
. The triangular shape of $\hat{M}$ shows that even the most diversified countries do not export just the most exclusive products, but even the common ones \cite{hausmann2007structure,hidalgo2007product,hidalgo2009building,PLoSONEFFQQ,tacchella2012new,tacchella2013economic, cristelli2013measuring,zaccaria2014taxonomy, Pugliese2014}. \\
The form of the distribution for the Fitness (Quality) ranking against country diversification (products ubiquity) depends strongly on the shape of $\hat{M}$. They are sparse distributions, due to the non-linearity of the algorithm of Fitness/Complexity, which cause even the peculiar shape, as shown in the Fig. \ref{fig:mah}(a, b): real data are blue points, while the cloud represent the frequency of the simulated data. 

\subsubsection*{Nestedness}
As already mentioned, the \coniglietti{triangularity} of the $\hat{M}$-matrix is a typical feature of biological mutualistic networks. Traditionally, in the biology literature the triangular form of the biadjacency matrix is measured by the \emph{nestedness}, i.e. how much a row (or a column) is a subset of the others. In the literature there is a great amount of different proposals for the nestedness definition, \cite{AlmeidaNetoetalNest,BastollaetalNest,AlmeidaNetoetalNest,MaritanetalNest,StaniczencoetalNest,MunozNest};  here we decide to use the NODF (\emph{Nestedness metric based on Overlap and Decreasing Fill}) presented  in \cite{AlmeidaNetoetalNest} because, according to us,  it is the most intuitive. 
Using the definition of Eqs. (\ref{eq:degree}), 
\begin{align*}
T_{cc'}^r\definition&\left\{
\begin{array}{c c}
k_c\neq k_{c'}& \dfrac{\sum_p m_{cp}\,m_{c'p}}{\text{Min}\{k_c, k_{c'}\}}\\
&\\
\text{otherwise} & 0\\
\end{array}
\right.;\\
T_{pp'}^c\definition&\left\{
\begin{array}{c c}
k_p\neq k_{p'}& \dfrac{\sum_c m_{cp}\,m_{cp'}}{\text{Min}\{k_p, k_{p'}\}}\\
&\\
\text{otherwise} & 0\\
\end{array}
\right..
\end{align*}
The total nestedness measure $\text{NODF}_t$ is then 
\beq\label{eq:NODFdef}
\text{NODF}_t\definition2\dfrac{\sum_{c<c'}^{N{_c}} T_{c\,c'}^r+\sum_{p<p'}^{N{_p}} T_{p\,p'}^c}{N_c(N_c-1)+N_p(N_p-1)}, 
\eneq
where $N_c$ and $N_p$ are the number respectively of countries and products.
Note that the final value of the nestedness gets contribution just in case the number of the non-zero elements of the two rows (or columns) considered is different (for that, the name ``decreasing fill" in the NODF acronym). Eventually, the final formula for the nestedness is the weighted sum of the contributions from rows and from columns, so it is possible to isolate the two different nestedness for rows (i.e. countries, $\text{NODF}_c$) and for columns (i.e. products, $\text{NODF}_p$):
\begin{equation}\label{eq:NODF2def}
\text{NODF}_c\definition2\dfrac{\sum_{c<c'} T_{c\,c'}^r}{N_c(N_c-1)};\qquad\qquad\text{NODF}_p\definition2\dfrac{\sum_{p<p'} T_{p\,p'}^c}{N_p(N_p-1)}. 
\end{equation}
Since usual matrices $\hat M$ are quite ``rectangular", i.e. the number of products is much greater then the number of countries, the total NODF \reff{eq:NODFdef} is biased by the contribution from the products. In fact, combining Eqs.\reff{eq:NODFdef} and \reff{eq:NODF2def}:
\begin{equation}\label{eq:NODFapprox}
\text{NODF}=\dfrac{N_c(N_c-1)\text{NODF}_c+N_p(N_p-1)\text{NODF}_p}{N_c(N_c-1)+N_p(N_p-1)}\overset{N_p\gg N_c}{\simeq}\text{NODF}_p.
\end{equation}

Because of the previous relation, for the analysis of the initial conditions we just compared the results for $\text{NODF}_c$ and $\text{NODF}_p$, while the effect of the previous approximation is shown in the comparison between the Table \ref{tab:nodft_2003} and the Table \ref{tab:nodfc_2003}.

\subsubsection*{Assortativity}
The assortativity parameter $r$ \cite{NewmanAssortativity} has been introduced in order to measure how much nodes tend to be linked by nodes with a similar degree.
More in details, $r$ takes values from -1 to 1, where -1 denotes a network perfectly disassortative, i.e. lowest degree nodes are linked to the highest degree ones, while 1 denotes a network perfectly assortative, i.e.  high degree vertices links with highest degree one; in terms of the adjacency matrix and node degrees, $r$ can be written as
\begin{equation*}
r\definition\dfrac{2M \big(2\sum_{c,\, p}k_c m_{cp}k_p-(\sum_ck_c^2+\sum_pk_p^2)^2\big)}{2M\big((\sum_ck_c^3+\sum_pk_p^3)-(\sum_ck_c^2+\sum_pk_p^2)^2\big)}.
\end{equation*}
For the previous discussions about the triangularity, one may expect that the value of assortativity is negative (since poorly diversified countries exports the most ubiquitous products), but with a low absolute value, say much less than 0.5, since high degree countries have low complexity products in their export basket, as well as more complex ones. This expectation is just partly satisfied, since the value of $r$ for the matrices observed is indeed negative, but its absolute value is quite large, say of the order of 0.6: at a second look at the Fig. \ref{fig:mah}(c), in effect, it is possible to see that the density of the export basket of most diversified countries moves toward the low degrees products, i.e. the most exclusive ones. 

\subsubsection*{Motifs for bipartite networks}
Some motifs for bipartite networks have been defined in the contest of biological mutualistic networks. One of the most used is the checkerboards score \cite{checkerboardsscores} , i.e. the number of patterns of $2\times2$ submatrices present in the biadjacency matrices as mutually-exclusive terms such as $\checker,\,\antichecker$\footnote{$\checker$ and $\antichecker$ are meant to be $2\times2$ patterns in the $M-$matrix where $\blacksquare$ and $\square$ represent respectively $m_{cp}=1$ and $m_{cp}=0$.}\footnote{Since we prefer absolute measures, thus not depending on the order of rows and column, in the total number of checkerboards patterns we consider the occurrence of both $\checker$ and $\antichecker$.}. The checkerboards score, in other words, measures the how much mutually exclusive are the choices made by different countries about the composition of their export basket. The total number of possible checkerboards patterns can be written as
\begin{equation*}
N_{\text{checkerboards}}\definition\dfrac{1}{2}\sum_{c,c'}\sum_{p,p'}m_{cp}(1-m_{cp'})(1-m_{c'p})m_{c'p'}.
\end{equation*}

Other several motifs for bipartite networks have been proposed in \cite{saraccoetal} in order to uncover the structural properties of the system at hand. Among others, we decide to focus on $V$ and $\Lambda$ motifs\footnote{The names $V$ ($\Lambda$) come from the fact that once the two layers of the bipartite network are rotated such that the countries layer is over the products one, the former (the latter) motifs look like \coniglietti{$V$}s (\coniglietti{$\Lambda}$s) between the layers.}. Respectively, the total number of $V-$motifs and $\Lambda-$motifs count the total co-occurrence of products in two different export baskets and of countries in the set of producer of two different goods; in term of the entries of the biadjacency matrix $\hat{M}$, they are defined as 
\begin{equation*}
\begin{split}
N_{V}\definition&\sum_{c<c'} \sum_p m_{c\,p}m_{c'\,p};\\
N_{\Lambda}\definition&\sum_{p<p'} \sum_c m_{c\,p}m_{c\,p'}.\\
\end{split}
\end{equation*}

\subsection*{Tuning parameters}
The model, described in details in section ``The model", has a four-dimensional parameter space. In order to determine which of the values of  $(\alpha, \,\beta,\,\gamma,\,k^0)$ are more compatible with observations we generated matrices for all possible values of the parameters and compared the results with the observed data. In particular, we focus on initial conditions with  $N_\text{roots}\leq25$, according to the conclusions of the following section. In the Tables \ref{tab:nodft_2003}, \ref{tab:nodfp_2003}, \ref{tab:nodfc_2003}, \ref{tab:r_2003}, \ref{tab:check_2003}, \ref{tab:V_2003}, \ref{tab:Lambda_2003}, simulated data have been generated for initial conditions with $N_\text{roots}=20$ and $P_0=0.3$. 

We observed that the parameter $k^0$ appears not to be influential on the measures  for $k^0\geq4$. It is worth noticing that 4 is even the average value of the minimum of the ubiquity for real matrices; effectively $k_p=4$ is the first value of the ubiquity for which the probability of a novelty exceed the probability of innovating. In the following we will present results with $k^0=4$; for every ($\alpha,\,\beta,\,\gamma$) configuration we generated 56 matrices.\\

 As the Tables \ref{tab:nodft_2003}, \ref{tab:nodfp_2003}, \ref{tab:nodfc_2003}, \ref{tab:r_2003}, \ref{tab:check_2003} , \ref{tab:Lambda_2003} show, once $\beta$ and $\gamma$ are fixed, i.e. the parameters of the second and the third choices, the value of the $\alpha$ parameter which are able to replicate real data are quite narrow around the value $\alpha=1.6$. Instead, for a fixed value of $\alpha$, the best results are in an area around the ``anti-diagonal" of the tables shown, so for decreasing $\beta$ values once $\gamma$ grows and vice versa.\\

The results for the number of $V-$motifs  in Table \ref{tab:V_2003} deserves a special treatment: in fact, while, the $\Lambda-$motifs are well reproduced, it happens quite rarely that the model is able to replicate the observed results. The meaning of the results reside in the definition of $V-$ and $\Lambda-$motifs: $V$s ($\Lambda$s) are the number of co-occurrence of countries (products) in the set of producers for every product (in the export set for every country). In effect, our algorithm is driven by a hierarchy imposed on the products, while no kind of structure is forced onto the countries set, so the constraints drive the total number of $\Lambda$s, while the number of $V$s is too ``free".

\subsection*{Tuning Initial Conditions}
In Table \ref{tab:IC} we compare different values of $N_\text{roots}$ from 10 to 40 (at steps of 5 roots) and different $P_0$, from 0.2 to 0.6 (at steps of 0.05); we fix the values of the parameter for the evolution dynamics among the best performing ones, according to the previous analyses, i.e. $\alpha=1.6,\,\beta=1,\,\gamma=0.6$ and $k^0=4$; for every initial conditions configuration we produced 50 simulated $M-$matrix. The results can be observed in the Fig. \ref{tab:IC}: for low values of $N_\text{roots}$, the discrepancy among different distributions relative to different values of $P_0$ is limited and very often cross the red line representing real data. Because of it, we mostly used low $N_\text{roots}\leq25$ , since they need less fine tuning on the $P_0$. Moreover, higher $P_0$ are less precise, especially for a higher number of roots: thus it seems that the mean number of roots per countries to make the algorithm start should be quite small, say $\sim6$.\\

We tried, imposing some offsets for every choice, even to make the algorithm start from no product, but the result are not satisfying since the usual measures on the matrix are not replicated.

\newpage 
\section*{Figure \& Table Legends}

\bigskip

\bigskip

\bigskip

\bigskip

\begin{landscape}

\begin{table}[htb] 
	\centering
		\includegraphics[width=1.5\textwidth]{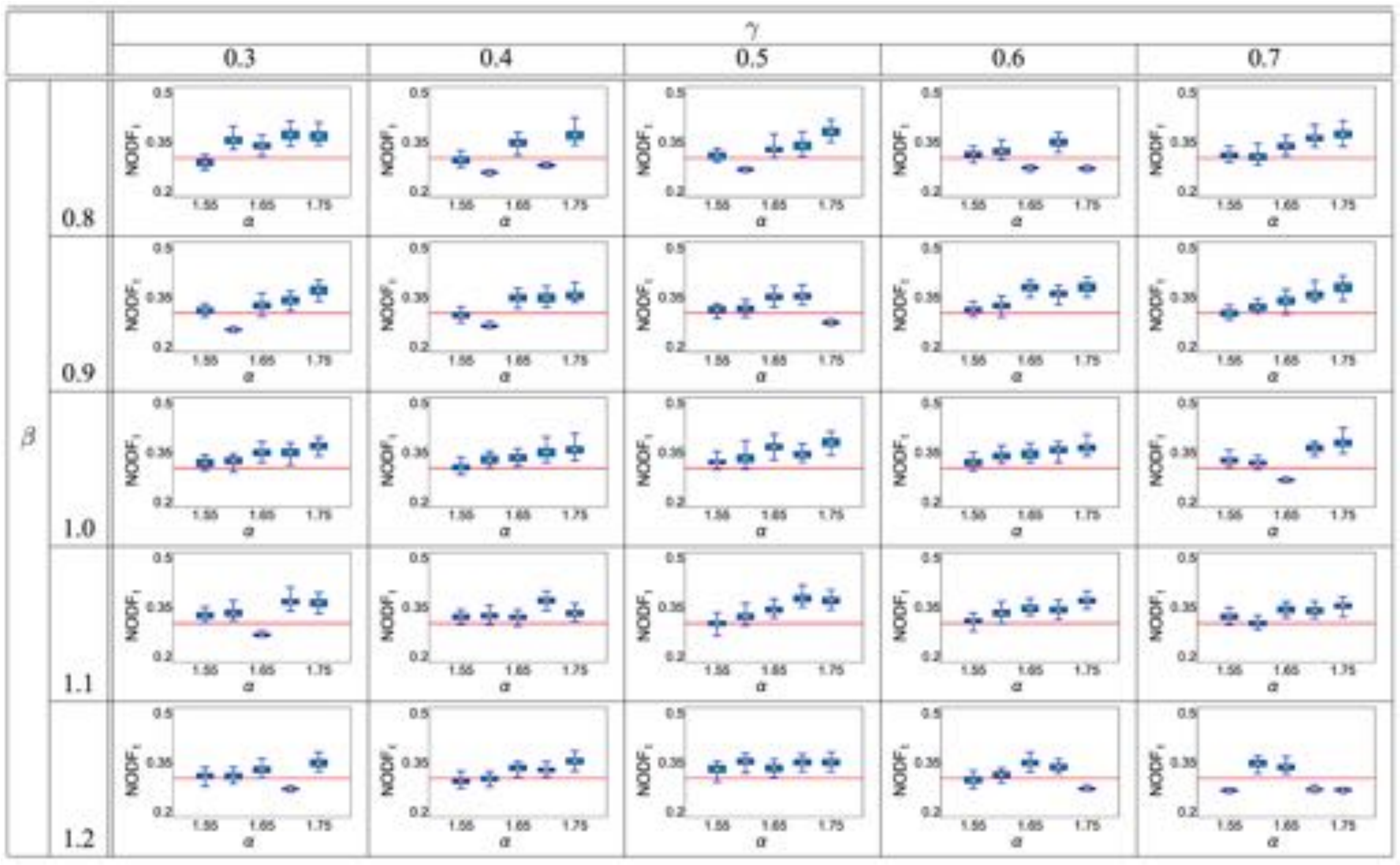}
\caption{Parameters space analysis: $\textbf{NODF}_{\textbf{t}}$. It is possible to observe the variation of the $\text{NODF}_t$ at the changing of the parameters $\alpha,\,\beta\,\gamma$; the parameter $k^0$ has been kept fixed to the value 4, since no variation in any of the measure analysed has been observed for greater values. The $\text{NODF}_t$ measured on the original matrix is represented as a red line. The best values of the parameter $\alpha$ are the lowest analysed, i.e. $\alpha\leq1.65$. Instead, the wider area of acceptance for the $\gamma$ parameters is for the central area of the table, i.e. $0.4\leq\gamma\leq0.6$, while $\beta$ is more or less non-influential on the the acceptance of the measure.}
\label{tab:nodft_2003}
\end{table}
\end{landscape}

\bigskip

\begin{landscape}
\begin{table}[htb]
	\centering
		\includegraphics[width=1.5\textwidth]{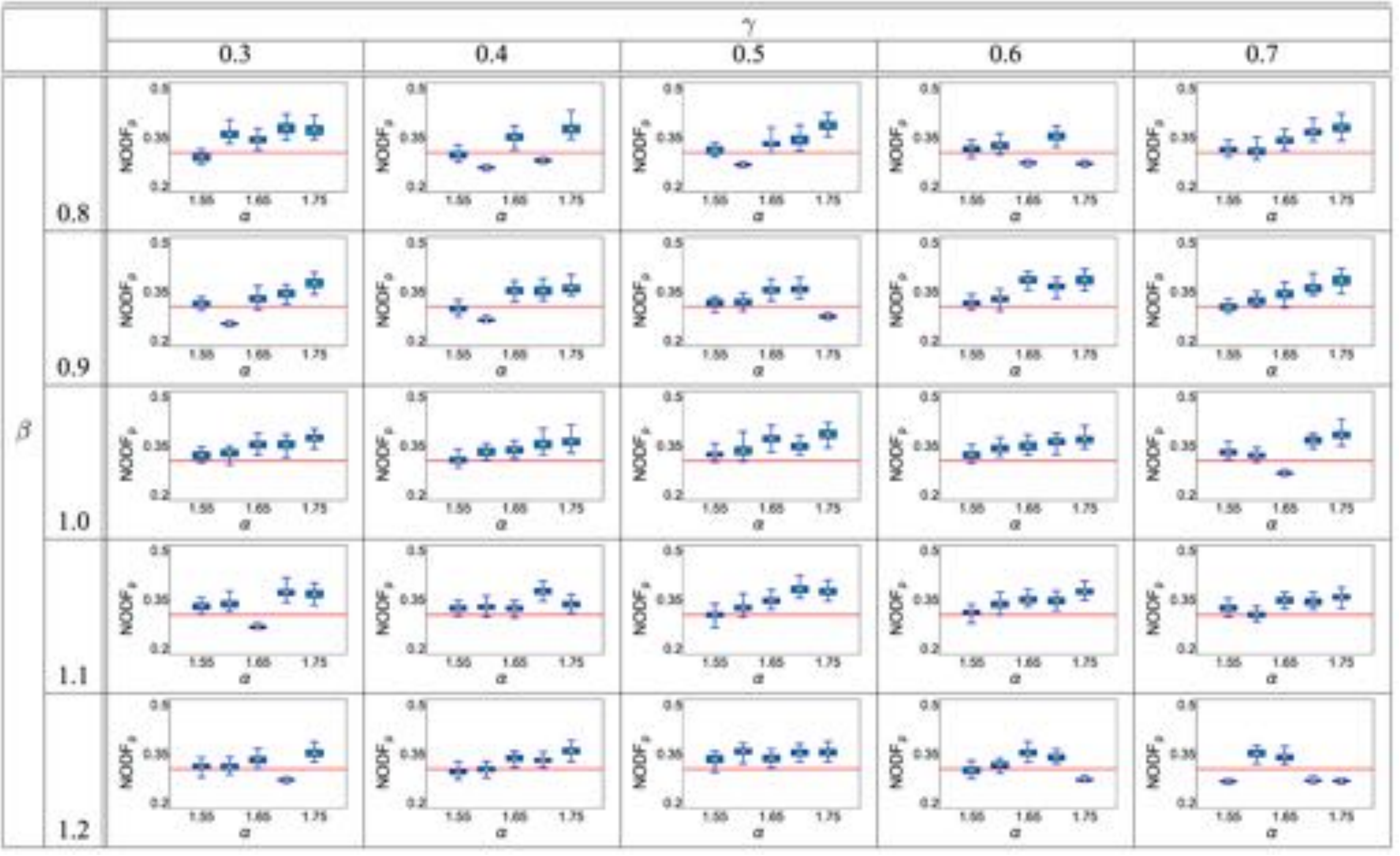}\caption{Parameters space analysis: $\textbf{NODF}_{\textbf{p}}$. It is possible to observe the variation of the $\text{NODF}_p$ at the changing of the parameters $\alpha,\,\beta\,\gamma$; the parameter $k^0$ has been kept fixed to the value 4, since no variation in any of the measure analysed has been observed for greater values. The $\text{NODF}_p$ measured on the original matrix is represented as a red line. The best values of the parameter $\alpha$ for reproducing $\text{NODF}_p$ are the lowest analysed, i.e. $\alpha\leq1.65$. Instead, the wider area of acceptance for the $\gamma$ parameter is for the central area of the table, i.e. $0.4\leq\gamma\leq0.6$, while $\beta$ is more or less non-influential on the the acceptance of the measure. Note that all the graphs presented here are completely overlapping with the one of the Table \ref{tab:nodft_2003}, because for the analysed network the approximation of Eq.(\ref{eq:NODFapprox}) holds. }
\label{tab:nodfp_2003}
\end{table}
\end{landscape}

\bigskip

\begin{landscape}
\begin{table}[htb]
	\centering
		\includegraphics[width=1.5\textwidth]{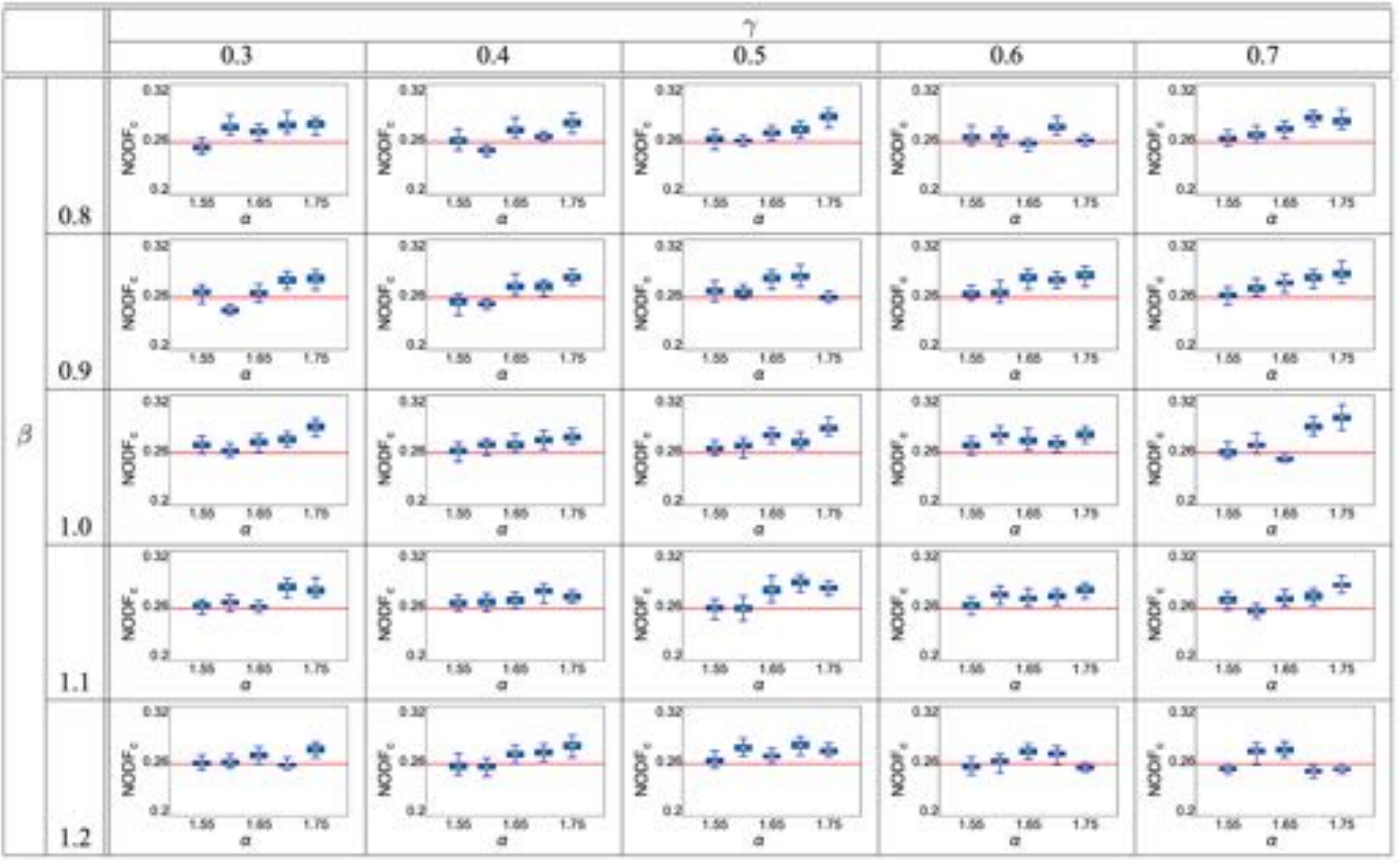}\caption{Parameters space analysis: $\textbf{NODF}_{\textbf{c}}$. It is possible to observe the variation of the $\text{NODF}_c$ at the changing of the parameters $\alpha,\,\beta\,\gamma$; the parameter $k^0$ has been kept fixed to the value 4, since no variation in any of the measure analysed has been observed for greater values. The $\text{NODF}_c$ measured on the original matrix is represented as a red line. The best values of the parameter $\alpha$ are the lowest analysed, i.e. $\alpha\leq1.65$, as in Table \ref{tab:nodfp_2003}. In replicating the measure of $\text{NODF}_c$ the value of $\beta$ and $\gamma$ are more non influential than in Table \ref{tab:nodfp_2003} and more or less all configuration $(\alpha, \beta, \gamma)$ are able to correctly replicate the real value.}
\label{tab:nodfc_2003}
\end{table}
\end{landscape}

\bigskip

\begin{landscape}
\begin{table}[htb]
	\centering
		\includegraphics[width=1.5\textwidth]{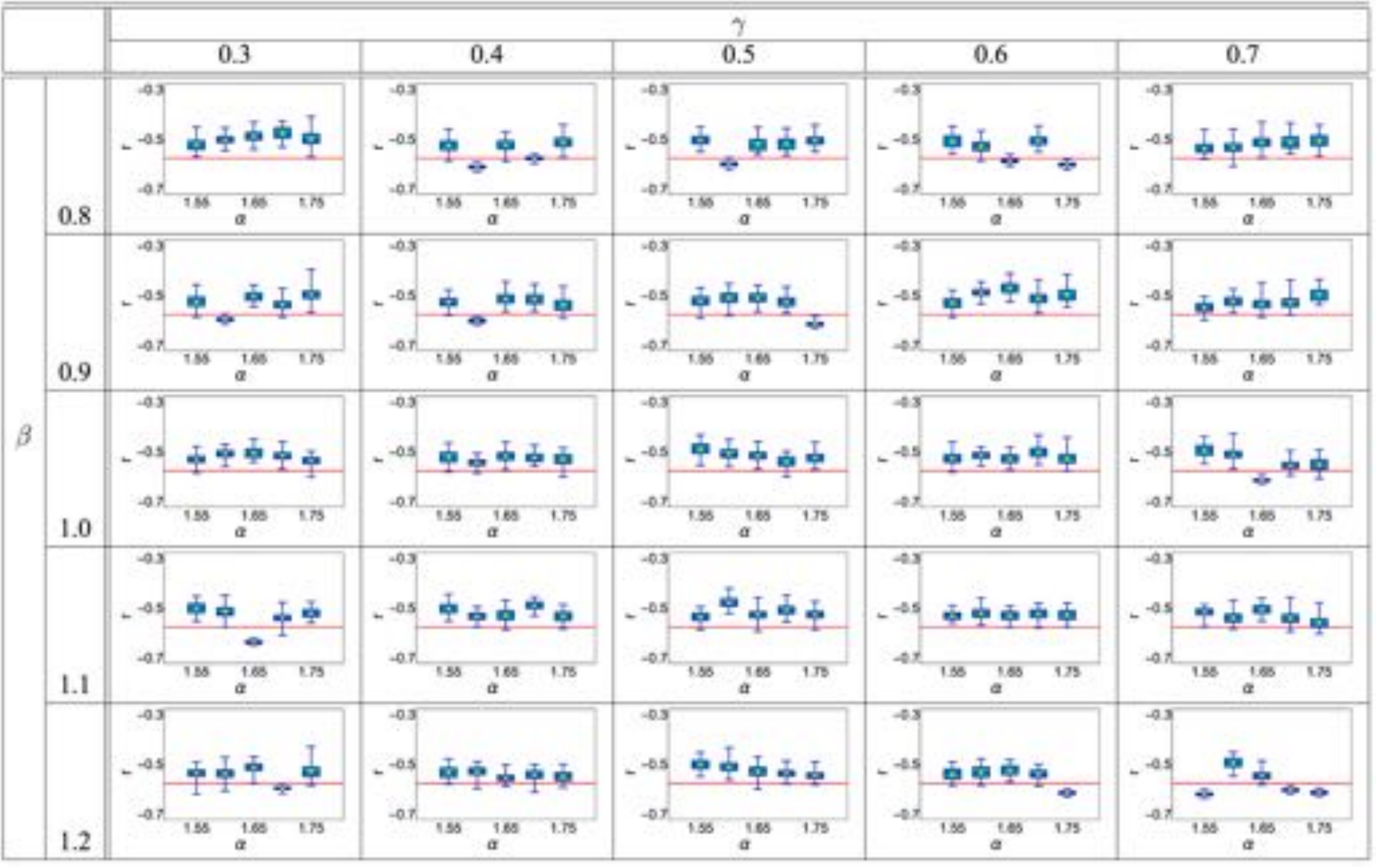}\caption{Parameters space analysis: \textbf{r}. It is possible to observe the variation of the $r$ at the changing of the parameters $\alpha,\,\beta\,\gamma$; the parameter $k^0$ has been kept fixed to the value 4, since no variation in any of the measure analysed has been observed for greater values. The r measured on the original matrix is represented as a red line. Instead, the wider area of acceptance for the $\beta$ and $\gamma$ parameters is around the ``anti-diagonal" of the table represent, so high value of $\gamma$ for low $\beta$ and vice versa, while $\alpha$ is always centred over 1.65. }
\label{tab:r_2003}
\end{table}
\end{landscape}

\bigskip

\begin{landscape}
\begin{table}[htb]
	\centering
		\includegraphics[width=1.5\textwidth]{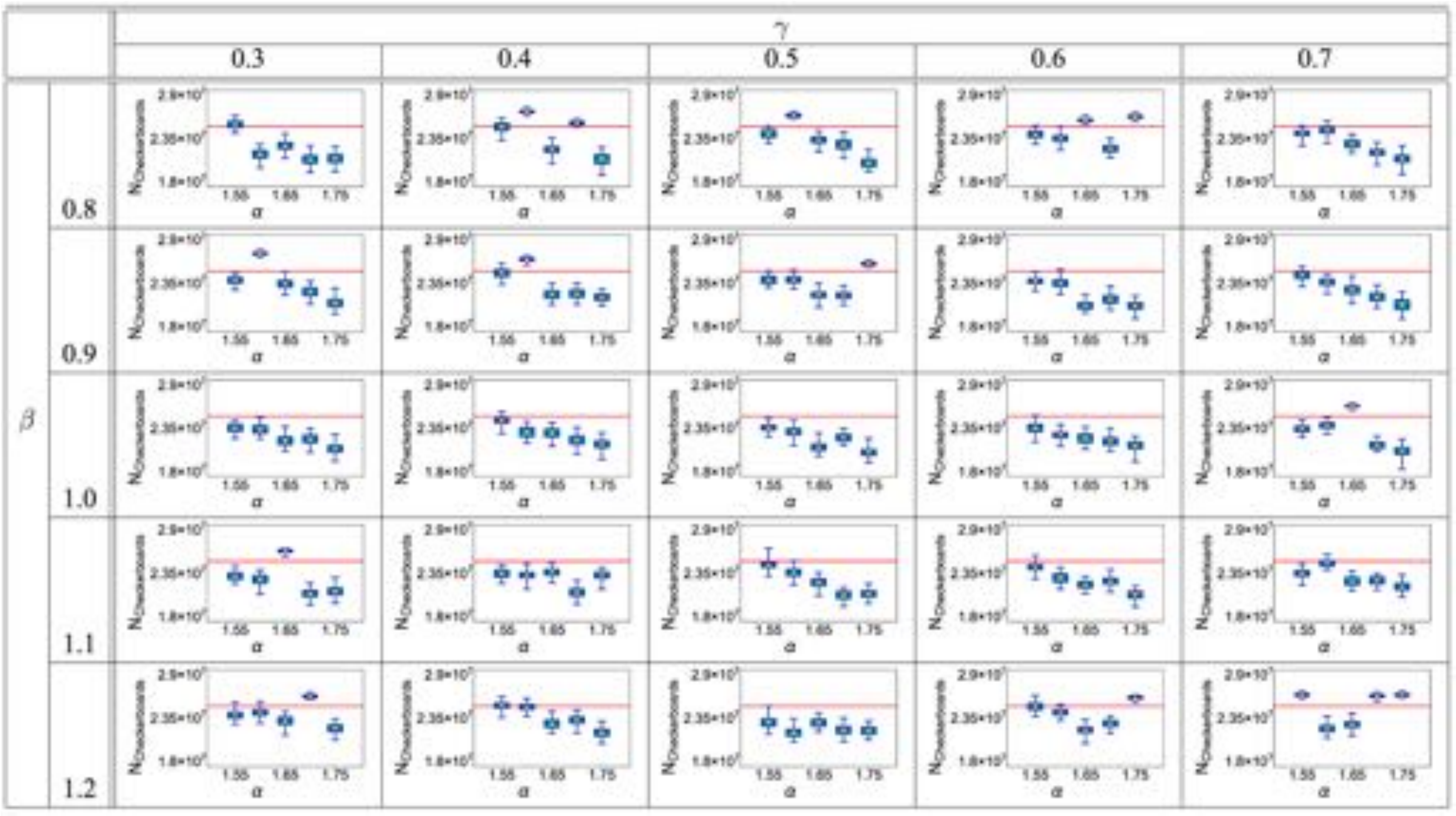}\caption{Parameters space analysis: $\textbf{N}_{\textbf{Checkerboards}}$. 
It is possible to observe the variation of the $N_\text{Checkerboards}$ at the changing of the parameters $\alpha,\,\beta\,\gamma$; the parameter $k^0$ has been kept fixed to the value 4, since no variation in any of the measure analysed has been observed for greater values. The $N_\text{Checkerboards}$ measured on the original matrix is represented as a red line. The wider area of acceptance for the $\beta$ and $\gamma$ parameters is similar to the one of the Table \ref{tab:r_2003}, while $\alpha$s performing better are a little bit lower, say $\alpha\leq1.65$.}
\label{tab:check_2003}
\end{table}
\end{landscape}

\bigskip

\begin{landscape}
\begin{table}[htb]
	\centering
		\includegraphics[width=1.5\textwidth]{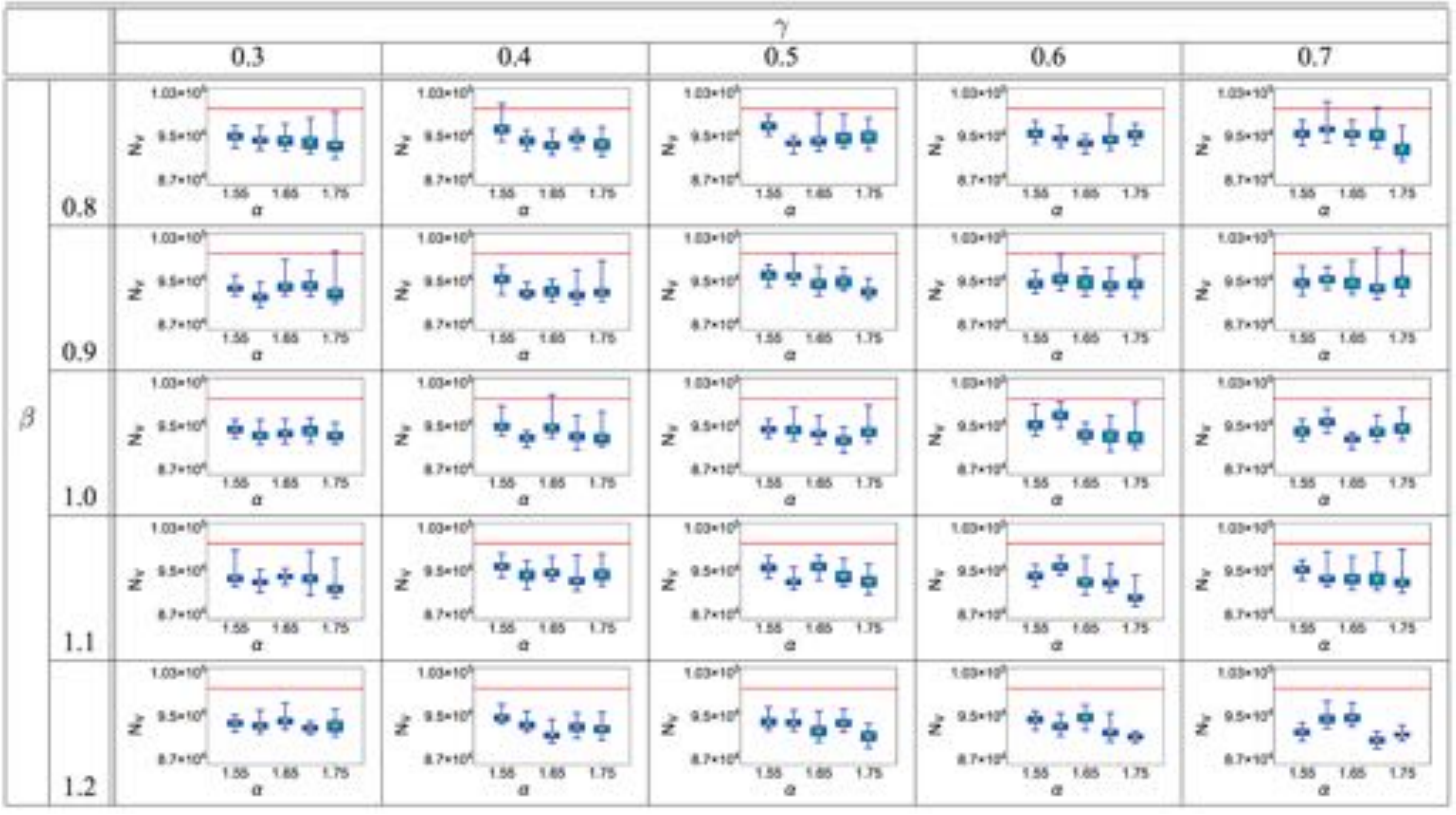}\caption{Parameters space analysis: $\textbf{N}_{\textbf{V}}$. 
It is possible to observe the variation of the $N_V$ at the changing of the parameters $\alpha,\,\beta\,\gamma$; the parameter $k^0$ has been kept fixed to the value 4, since no variation in any of the measure analysed has been observed for greater values. As it is possible to see in the present table, our model is not able to capture the number of $V-$motifs in the network for more or less none of the parameters analysed. This phenomenon is due to the fact that the model evolution is based on a hierarchical structure for products (the products network) that is not present for the countries: in effect, as Table \ref{tab:Lambda_2003} shows, there is much more agreement with the original data for the $\Lambda-$motifs.}
\label{tab:V_2003}

\end{table}
\end{landscape}

\bigskip

\begin{landscape}
\begin{table}[htb]
	\centering
		\includegraphics[width=1.5\textwidth]{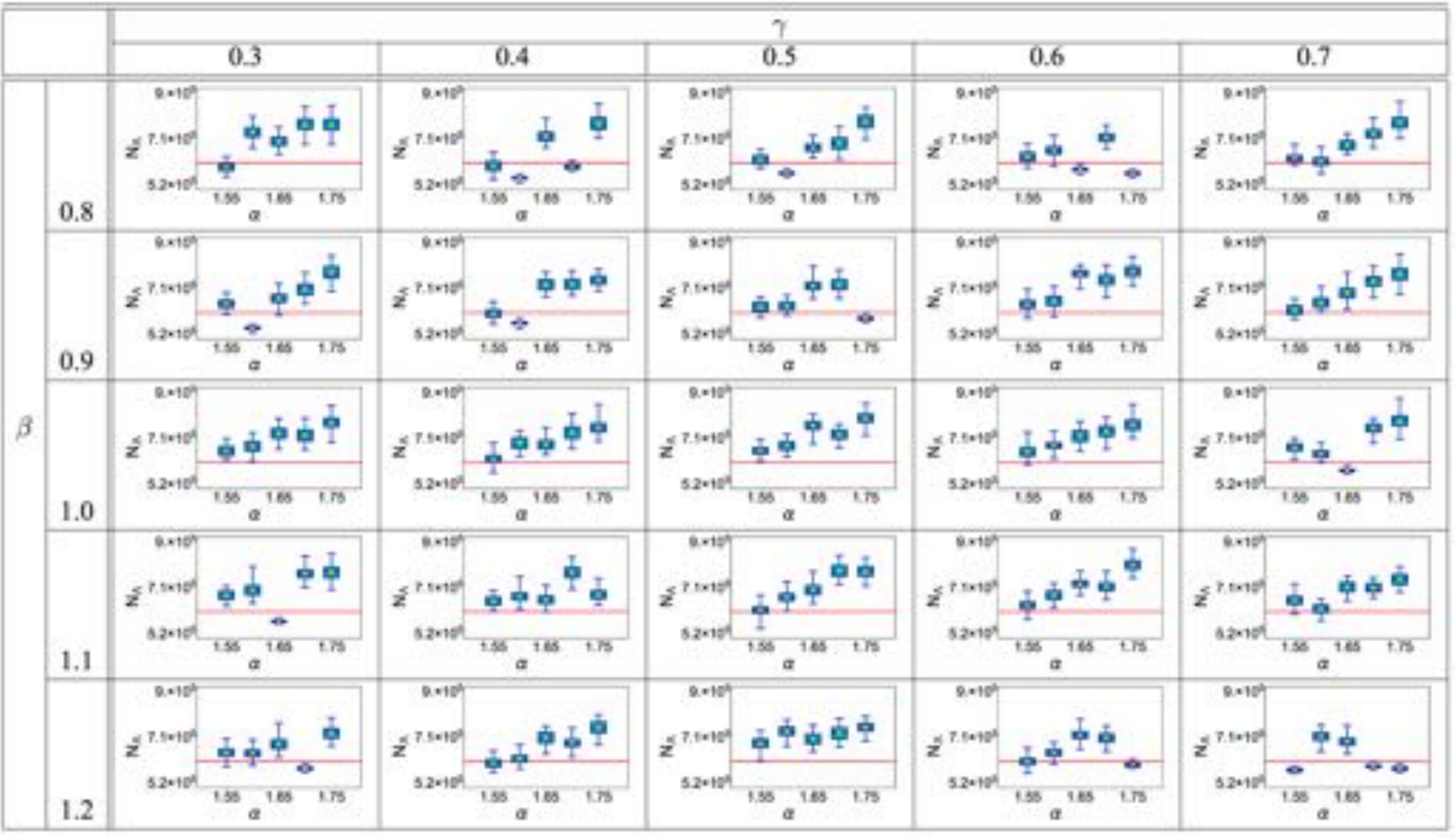}
\caption{Parameters space analysis: $\textbf{N}_\bold{\Lambda}$. 
It is possible to observe the variation of the $N_{\Lambda}$ at the changing of the parameters $\alpha,\,\beta\,\gamma$; the parameter $k^0$ has been kept fixed to the value 4, since no variation in any of the measure analysed has been observed for greater values. Respect to the Table \ref{tab:V_2003} we find a better agreement in reproducing $\Lambda-$motifs: this fact is probably due to base the evolution of the model on a (evolving) structure for products, which keeps trace in the total number of $\Lambda-$motif, i.e. the number of co-occurrence of 2 different products in the exports baskets. As in the previous tables, the best results are obtained for low value of $\alpha$ and the area along the ``anti-diagonal" of the presented table.}

\label{tab:Lambda_2003}
\end{table}
\end{landscape}

\bigskip

\begin{landscape}
\begin{table}[htb]
	\centering
		\includegraphics[width=1.35\textwidth]{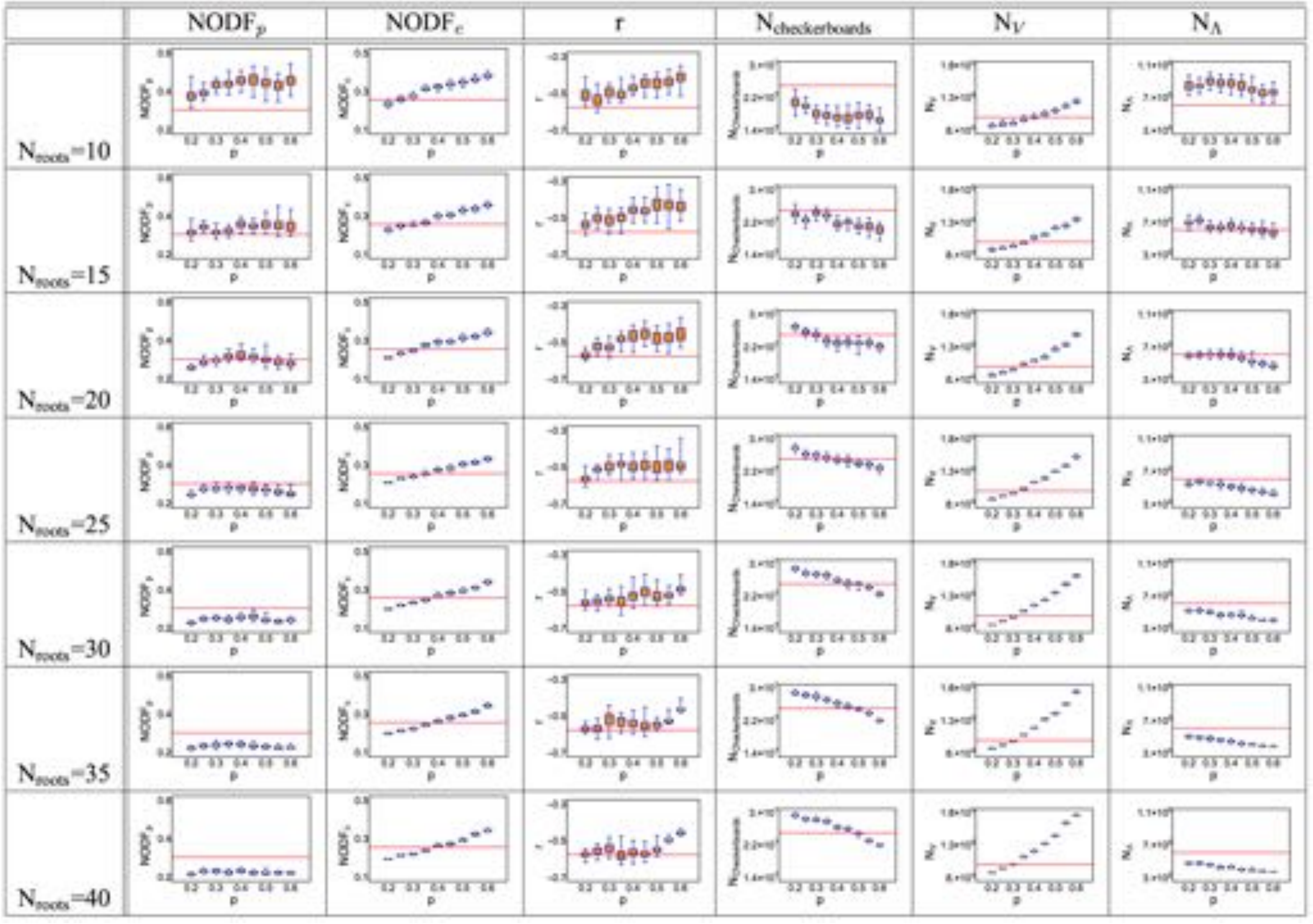}
		\caption{Initial Conditions analysis: in the Table \ref{tab:IC} it is possible to observe how the measurements vary in changing the initial condition (for fixed parameter: $\alpha=1.6,\,\beta=1,\,\gamma=0.6,\,k^0=4$) as compared with the values from the real matrix (the red line). Every boxplot contains the distribution of 56 simulations. Initial conditions are performed assigning $N_\text{roots}$ initial products roots, each with probability $P_0$. As it is possible to see, the best agreement for most of the measures used is for $10\leq N_{\text{roots}}\leq25$ and $P_0\leq0.4$.}
\label{tab:IC}
\end{table}
\end{landscape}

\end{document}